\documentstyle[epsfig,twocolumn]{mn}

\title{Reprocessed emission line profiles from dense clouds in geometrically 
thick accretion engines}

\author[Sean A. Hartnoll and Eric G. Blackman]
{{Sean A. Hartnoll$^{1,2}$ and Eric G. Blackman$^1$} \\
{$^1$ Department of Physics and Astronomy, University of Rochester,
Rochester NY 14627, USA} \\
{$^2$ St. John's College, Cambridge University, Cambridge CB2 1TP, UK}}

\date{Submitted version}

\pubyear{2001}

\def\be{\begin{equation}}
\def\ee{\end{equation}}
\def\vn{{\bf \hat{n}}}
\def\vr{{\bf r}}
\def\vro{{\bf r_0}}
\def\vrobs{{\bf r_{obs}}}
\def\vc {{\bf c}}
\begin{document}

\maketitle

\begin{abstract}
The central engines of active galactic nuclei (AGN) 
contain cold, dense material 
as well as hot X-ray emitting gas. The standard paradigm for the engine 
geometry is a cold thin disc sandwiched between hot X-ray coronae.
Strong support for this geometry in Seyferts comes from the study of 
fluorescent iron line profiles, 
although the evidence is not ubiquitously air tight.
The thin disc model of line profiles in AGN and in X-ray binaries
should be bench marked against  
other plausible possibilities. One proposed alternative 
is an engine consisting of 
dense clouds embedded in an optically thin, geometrically thick X-ray 
emitting engine. This model is further 
motivated by studies of geometrically thick engines
such as advection dominated accretion flows (ADAFs).  
Here we compute the reprocessed iron line profiles from dense
clouds embedded in geometrically thick, optically thin X-ray emitting 
discs near a Schwarzchild black hole. 
We consider a range of cloud distributions and disc solutions,
including ADAFs, pure radial infall, and bipolar outflows.
We find that such models can reproduce line profiles similar to those from 
geometrically thin, optically thick discs and might help alleviate some of the 
problems encountered from the latter.
\end{abstract}


\section{Introduction}

Active galactic nuclei (AGN) and some X-ray binaries are thought to be
powered by accretion onto massive black holes (e.g. Rees 1984; Tanaka 2000).
Observations suggest that 
cold matter in the accretion engine reprocesses the primary
X-ray continuum (e.g. George and Fabian 1991). In particular, 
a fluorescent iron K$\alpha$ emission line profile is imprinted on the
spectrum near 6.4 keV. The iron line profile is 
shaped by Doppler and gravitational frequency shifts (Fabian et al. 1989), 
and hence contains information about geometry and dynamics in the innermost
regions of the accretion engine
(see Fabian et al. 2000 for a comprehensive review).

Most modelling of iron line profiles has been carried out
within the paradigm of geometrically thin, optically thick discs.
The initial flat disc model of Fabian et al. (1989)
was generalised to Kerr black holes by Laor (1991).
More recently, Pariev and Bromley (1998)
have considered effects of a small disc thickness and turbulence.

Thin disc line profiles successfully reproduce many features in the
experimental data (e.g. Tanaka et al. 1995; Nandra et al. 1997).
However there are some observations which remain 
difficult to understand within this paradigm. For example, 
a preference for low inclinations,
even amongst Seyfert 2s (Turner et al. 1998),
disagreements between the predicted inclination from iron line
modelling and the predicted inclination from other
phenomena (Nishiura et al. 1998; Sulentic et al. 1998a; Wang et al. 1999), 
and problems with the observed reprocessed fraction being too low
or not correlated with line width
(e.g. Lee et al. 2000; Chiang et al. 2000;
Weaver 2000; Sulentic et al. 1998b).

Such puzzles may be resolved within the paradigm of thin
discs, for instance by introducing additional emission components from
an obscuring torus (Weaver and Reynolds 1998) or invoking
flared or warped discs in which the inner and outer regions are not
aligned (Blackman 1999; Hartnoll and Blackman 2000). 
In any case, alternative 
plausible geometries still merit exploration to provide proper
diagnostics for comparison.

A particularly interesting alternative extends from 
work on the survival of cold clouds capable of reprocessing X-ray flux 
in the AGN engines 
(e.g. Ferland and Rees 1988; Guilbert and Rees 1988; Celotti, Fabian
and Rees 1992). 
Physical constraints on the size of the dense clouds
have been studied by Kuncic et al. (1996). They found
a sizeable parameter regime in which cool, dense, small clouds
can coexist with a hot medium in a dynamical equilibrium.
Though X-ray continuum features of the embedded cloud model have been 
studied by several authors (Sivron and Tsuruta 1993; Bond and Matsuoka 1993;
Nandra and George 1994; Collin-Souffrin et al. 1996),
detailed calculations of reprocessed line profiles have
not been carried out.   
Karas et al. (2000) have studied line emission from 
reprocessing by a nearly spherical shell of clouds at a specific radius
that results from a standard thin disc being disrupted due to instabilities.
They give resulting line profiles from illumination by a central source.
This is a different scenario from what we will consider here.
Other related calculations of note include Cao and Zhang (1991),
who calculated line profiles for radially infalling gas.

In this paper we calculate the (iron) line profiles 
that result from various distributions of small, cold clouds 
embedded in an optically
thin, geometrically thick accretion disc or corona.
As well as illumination from a central source,
we include an illumination by an `X-ray bath' of
radiation; each cloud is taken to be small enough to be 
illuminated isotropically at their positional location
by this bath component. 
This two component illumination is motivated from 
considering geometrical thick models such as,
for example, advection dominated accretion flows 
(ADAFs) (e.g. Rees et al. 1982; 
Narayan and Yi 1994; see Narayan et al. 1998 for a review), 
which can be interpreted to 
have a synchrotron + Bremsstrahlung centrally dominated
illumination (Mahadevan 1997) and a Bremsstrahlung X-ray bath.
Depending on the accretion rate of the disc, the synchrotron
or Bremsstrahlung will dominate.
ADAFs predict luminosities significantly less than the Eddington value $L_E$, 
and may just be applicable to Seyferts, with
typical luminosities $L_E/100$. 
 
We will use several flow solutions 
to prescribe a velocity field for the clouds;
standard full ADAF solutions, pure radial infall, and bipolar outflows.
The profiles calculated
may also therefore be relevant for the dynamics of convection
dominated accretion flows (CDAFs) (Quataert and Gruzinov 2000; Narayan
et al. 2000), Advection Dominated Inflow Outflow Models 
(ADIOS) (Blandford and Begelman 1999) or other thick disc models.
The microphysics of the direct X-ray line emission  
for ADAFs have been
studied in non-relativistic flows (e.g. Narayan and Raymond 1999; 
Perna et al. 2000), but not the presence of cold reprocessing
matter, nor the associated relativistic iron flourescence 
line profiles. Like Karas et al. (2000), we 
take the clouds to be optically thick, and allow the clouds to
shadow emission from other clouds.

Some profiles for relativistic outflows have been calculated
by Wang et al. (2000). These outflows are different from those considered
here in that they are pencil beams along the spin axis and not
radial outflow in a geometrically thick disc.
The spherical geometry, and covering by clouds, can avoid the large blue
shifted component that has not been observed in iron lines and
that is characteristic of outflows.

It has been suggested (e.g. Nandra 2000) that a geometrically
thick distribution of clouds would 
result in more blue-shifted emission than is observed.
The data from a variety of sources are also not inconsistent with 
a somewhat larger blue wing (Nandra et al. 1997).
But we find that because models of geometrically thick discs can generally
have smaller bulk velocities in the central regions than
the flat disc keplerian velocities, smaller 
Doppler blue shifts result, and gravitational
redshift becomes the dominant spectral feature.
For these reasons also, the line profiles for thick discs will
in general be less inclination dependent than those from thin discs.
This will be discussed in more detail below. 
It may help explain
problems encountered with inclinations in thin disc modelling.
When the velocities are faster, the profiles are very similar
to the standard profiles obtained from thin disc modelling.

In section 2 we describe the disc model of dense clouds in
geometrically thick discs. Section 3
contains the calculation of iron line profiles
from this model. Section 4 presents the resulting line profiles
and a discussion of the main features.
Section 5 is the conclusion.

\section{The embedded cloud model: dense clouds and thick discs}

The model consists of small optically thick
clouds embedded in, and hence following, the fluid flow in an
otherwise optically thin
disc. Although the clouds may have a short lifetime, the continuous
creation and destruction of clouds allows the definition
of an effective number density of clouds at each point, $n_c(\vr)$.
We consider these clouds to be both bathed in an locally isotropic radiative
flux (e.g. Bremsstrahlung) and to be
illuminated by a central source (e.g. synchrotron dominated).
The clouds are assumed to be small compared to disc length scales 
(see Celotti et al. 1992 and Kuncic et al. 1996 for partial justification).

Calculating the profile then involves an
optical depth calculation; we integrate the number density times
the cross section along the photon trajectories to obtain
a probability of obscuration. Contributions are integrated across
the whole disc to obtain a profile, as is done for
the usual thin disc,
except that now we are integrating over three dimensions.

We consider several different solutions to the accretion flow
involving optically thin, geometrically thick engines:
self-similar ADAFs (Narayan and Yi 1995;
Perna et al. 2000),
bipolar outflows (Xu and Chen 1997; Perna et al. 2000),
and radial infall (appendix B of Narayan and Yi 1995, similar to Bondi 1952).

\subsection{self-similar ADAF}

Self-similar accretion solutions are described in Narayan and Yi (1995).
The approximate analytic solutions, given in Perna et al. (2000), for
the density ($\rho$), velocities ($v_r, v_{\theta}, v_{\phi}$), and
sound speed ($c_s$) are 
\be\label{eq:one}
\rho = \rho_0 r^{-\frac{3}{2}} [c_s^2(\theta)]^{-2.25} ,
\ee
\be\label{eq:two}
v_r = - \frac{3 \alpha}{5 + 2 \epsilon} r \Omega_K(r) \sin^2\theta ,
\ee
\be
v_{\theta} = 0 ,
\ee
\be\label{eq:three}
v_{\phi} = \left( \frac{2 \epsilon}{5 + 2\epsilon} \right)^{\frac{1}{2}}
r \Omega_K(r) \sin\theta ,
\ee
\be\label{eq:four}
c_s = r \Omega_K(r) \left[\frac{2}{5}-
\left(\frac{2}{5} - \frac{2}{5+2\epsilon}\right)\sin^2\theta
\right]^{\frac{1}{2}} ,
\ee
where $\Omega_K(r)$ is the
Keplerian angular velocity, $\Omega_K(r) = r^{-\frac{3}{2}}$ in our units
and we are using standard spherical polar coordinates, $(r,\theta,\phi)$.

There are two free parameters, the Sakura-Sunyaev viscosity parameter,
$\alpha$, and a thermodynamic parameter,
\be
\epsilon = \left( \frac{5/3 - \gamma}{\gamma - 1} \right) ,
\ee
with $\gamma$ the ratio of specific heats. We will consider
$\alpha = 0.35$ and $\epsilon = 10, 0.1$. Solutions are taken
to be in the fully advection dominated case. Note that:
\begin{enumerate}
\item Lower $\epsilon$ gives lower angular speeds.

\item For $\epsilon = 10$ most of the matter is close to the equatorial plane,
whilst for $\epsilon = 0.1$ the matter is essentially uniformly distributed
in $\theta$.

\item For larger values of $\alpha$, such the $\alpha=0.35$ we consider,
$v_r$ is very similar to $v_{\phi}$. For smaller $\alpha$, the radial
motion becomes less important.

\item $c_s(\theta)$ is smaller on the equatorial plane
and is smaller for larger $\epsilon$. This will mean
that, at fixed radius,
the clouds are larger on the equatorial plane and larger
in high $\epsilon$ cases.
\end{enumerate}

\subsection{Radial infall}

The self-similar forms of the previous section admit a radially
infalling solution (Narayan and Yi 1995, it is essentially Bondi
accretion with viscosity.) We take the radial infall to be at the
Keplerian velocity, $r \Omega_K(r)$, which implies that the
sound speed is $c_s = \frac{r \Omega_K(r)}{\sqrt{5}}$. This
solution corresponds to a small $\epsilon$, and so represents a
geometrically thick, optically thin, advection dominated flow.
The density is taken to be constant in $\theta$, so
$\rho = \rho_0 r^{-\frac{3}{2}}$.

\subsection{Bipolar outflow}

Outflow solutions have been considered by several
authors (e.g. Xu and Chen 1997; Quartaert and Narayan 1999;
Blandford and Begelman 1999).
Perna et al. (2000) give approximate analytic expressions.
We use their solutions with density independent of angle
and with winds of $p=\frac{1}{2}$, where $p$ is
defined so that $\dot{M} \propto r^p$. The equations then become
\be\label{eq:f}
\rho = \rho_0 r^{-\frac{3}{2}} ,
\ee
\be
v_r = \frac{3\alpha}{5+2\epsilon} r \Omega_K(r) [\cos2\theta + \cos^2\theta] ,
\ee
\be
v_{\theta} = \frac{3\alpha}{5+2\epsilon} \frac{r \Omega_K(r)}{4} \sin2\theta ,
\ee
\be
v_{\phi} = \left( \frac{2 \epsilon}{5 + 2\epsilon} \right)^{\frac{1}{2}}
r \Omega_K(r) \sin\theta ,
\ee
\be\label{eq:s}
c_s = \frac{r \Omega_K(r)}{\sqrt{5+2\epsilon}}\sqrt{2-\epsilon \cos^2\theta} .
\ee

\section{Calculation of the line profile}

\subsection{Total flux}

The total flux observed at frequency $\nu$ is given by
\be
F(\nu) = \int \delta (\nu - \nu_0(\vr)) f(\vr,\nu) d^3r ,
\ee
where $\nu_0(\vr)$ is the observed frequency of flux emitted
from $\vr$, calculated below, and $I(\vr,\nu)$ is given by
\be
f(\vr,\nu) = \left[ \frac{\nu}{\nu_e} \right]^3
f_{out}(\vr,\vc(\vr)) V(\vr) ,
\ee
where $\left[\frac{\nu}{\nu_e}\right]^3$ is the relativistic correction
to the flux, with the frequency of emission,
$\nu_e = 6.4 {\mathrm{keV}}$ for the iron line.
$f_{out}(\vr,\vc(\vr))$ is the flux reprocessed and re-emitted
by clouds at $\vr$ in the direction $\vc(\vr)$,
defined such that the photons
reach the observer (calculated below). Finally, $V(\vr)$
is the {\it visibility function} which gives the probability
of flux from $\vr$ reaching the observer without being intercepted
by an optically thick cloud. It is given by
\be
V(\vr) = \max (1 - P(\vr), 0) ,
\ee
with the probability of interception
\be\label{eq:int}
P(\vro) = \int_a^b n_c(\vr(\lambda)) A(\vr(\lambda),\frac{d\vr}{d\lambda})
\left| \frac{d \vr}{d \lambda} \right| d \lambda ,
\ee
where $\vr(a) = \vro$, the position of emission, and
$\vr(b) = \vrobs$, the position of the observer, and $\lambda$
parametrises the path of the photon. $n_c(\vr)$ is the number density
of clouds in the disc and $A(\vr,\frac{d\vr}{d\lambda})$ is the cross
sectional area of the clouds at $\vr$, as seen by a photon travelling
in direction $\frac{d\vr}{d\lambda}$. These quantities are discussed below.
The visibility function must also include the possibility of  photons
falling into the event horizon.

The dimensions of the disc are taken to be
$6 \leq r \leq 300$, $0 \leq \theta \leq \pi$ and $0 \leq \phi < 2 \pi$,
although there will not necessarily be a significant number density of clouds
throughout this whole region (because the number density of particles in
the disc itself gets smaller for larger $r$ and, depending on the
thermodynamic parameter $\epsilon$, gets smaller further
form the equatorial plane).
Also, the volume element has a general relativistic correction, so
\be
d^3r = \frac{dr}{\sqrt{1-\frac{2}{r}}}r^2 \sin\theta d\theta d\phi .
\ee

\subsection{Photon paths}

We use paths in the Schwarzchild metric, approximated by a first
order perturbation. We take units with $c = G = 1$ and
take the mass of the central object to be $M = 1$. Note that in
this section $\vr$ is the position of the photon travelling
from the cloud to the observer, and $\vro$ is the position of the
emitting cloud in the disc. We write $r_0 = \mid \vro \mid$ and
$r = \mid \vr \mid$.

First order perturbation of the Schwarzchild null geodesic equations gives
the photon path to be
\be
\frac{1}{r(\Phi)} = \frac{\sin\Phi}{2 R} +
\frac{1 + C\cos\Phi + \cos^2\Phi}{4 R^2} +
{\mathcal{O}} (\frac{1}{R^3}) .
\ee
Where $R$ and $C$ are constants, $r(\Phi)$ is the radial
distance from the origin at the central object and $\Phi$
is a polar angle on the plane of the trajectory. The
trajectory plane is defined by
the origin, the direction to the observer, $\vn$, and the initial position
vector $\vro$.

The boundary conditions are that the photon starts at
$(r_0, \Phi_0)$ and ends at $(\infty, \Phi_{obs})$, written
in $(r,\Phi)$ coordinates on the trajectory plane.
The boundary conditions are simplified if we choose the axis
on the trajectory plane that defines $\Phi$ to be in
the direction towards the observer, $\vn$, so that $\Phi_{obs} = 0$.
The constants are then given by
\be
C = -2 ,
\ee
\be
R^{\pm} = \frac{r_0 \sin\Phi_0}{4} \pm \frac{r_0}{4}
\sqrt{\sin^2\Phi_0 + \frac{4}{r_0} (1-\cos\Phi_0)^2 } .
\ee
Note that the minus sign case results in $R$ negative, which is not physical,
so we take the positive solution. Selecting this single path is effectively
not considering orbits that loop around the central hole one or more times,
consistent with the perturbative approach.

With this choice of axis, $\Phi_0$ is given by
\be
\cos\Phi_0 = \frac{\vro \cdot \vn}{r_0} .
\ee
In principle, there could be a problem inverting this expression due to
ambiguities in the definition of the inverse cosine. However,
these are avoided if we take the other axis on the trajectory plane
to be $\frac{\vro - (\vro \cdot \vn) \vn}{r_0}$. This vector,
orthogonal to $\vn$, is such that $\vro$ has positive component
in this direction. Therefore, we may take $\Phi_0 \in \left[ 0,\pi \right]$
with no ambiguities and the trajectory is given by
\be
\vr(\Phi) = r(\Phi) \left[ \cos\Phi \vn +
\sin\Phi \frac{\vro - (\vro \cdot \vn) \vn}{r_0} \right] .
\ee
This gives the probability of interception as
\be
P(\vro) = \int_o^{cos^{-1}\frac{\vro \cdot \vn}{r_0}}
n(\vr(\Phi)) A(\vr(\Phi),-\frac{d\vr}{d\Phi})
\left| \frac{d \vr}{d \Phi} \right| d\Phi ,
\ee
and the angle of emission is given by
\be
\vc(\vro) = - \frac{d \vr}{d\Phi} (\Phi = \Phi_0) .
\ee
The minus sign in the last two equations results because the photon
travels in the direction of decreasing $\Phi$.

\subsection{Frequency shifts}

The observed frequency is given in terms of the emitted frequency by
\be\label{eq:shift}
\nu_0 = \nu_e \frac{(p_{\mu} U^{\mu})_{obs}}{(p_{\mu} U^{\mu})_{em}} ,
\ee
where $p^{\mu}$ is the photon 4-momentum and $U^{\mu}$
is the 4-velocity of the emitter or observer. We employ
general relativity in the Schwarzchild metric, $g_{\mu \nu}$.
Write $p^{\mu} = (p^t,{\mathbf{p}})^{\mu}$, and
$U^{\mu} = \frac{dt}{d\tau} (1,{\mathbf{u}})^{\mu}$.
The time component of photon 4-momentum in the Schwarzchild metric is
\be
p^{t} = \frac{E}{1-\frac{2}{r}} ,
\ee
Where E is constant along the photon trajectory.
So requiring the photon 4-momentum to be null, $g_{\mu\nu}p^{\mu}p^{\nu} = 0$,
gives ${\mathbf{p}}^2 = g_{ij} p^i p^j  = \frac{E}{1-\frac{2}{r}}$ to
be the appropriate normalisation for the spatial components.
In the Schwarzchild metric, we have
\be
\frac{dt}{d\tau} = \frac{1}{\sqrt{1-\frac{2}{r}-\frac{u_r^2}{1-\frac{2}{r}}-
u_{\theta}^2 - u_{\phi}^2}} .
\ee
By definition,
\be
p_{\mu} U^{\mu} = g_{\mu \nu} p^{\mu} U^{\nu} ,
\ee
where the observer has ${\mathbf{u}}_{obs} = 0$  and
$r_{obs} \to \infty$. The emitter has ${\mathbf{p}}_{em} \propto \vc$ and
$r_{em} = r_0$. Putting all this into (\ref{eq:shift}) gives
\be
\nu_0(\vro) = \frac{\nu_e \sqrt{1-\frac{2}{r_0}
-\frac{u_r^2(\vro)}{1-\frac{2}{r_0}} -u_{\theta}^2(\vro) -u_{\phi}^2(\vro)}}
{1 - \frac{c_r(\vro) u_r(\vro)}{1-\frac{2}{r_0}}
- c_{\theta}(\vro) u_{\theta}(\vro) - c_{\phi}(\vro) u_{\phi}(\vro)} .
\ee
Here ${\mathbf{u}}(\vro)$ is the velocity of the clouds at $\vro$.
We take the velocity of clouds to trace the velocity of the fluid
in the disc. This is given by one of the disc motions described
in the previous section.

\subsection{Distribution and sizes of clouds}

For simplicity, we will assume that all of the clouds are spherical.
The cross section is then independent of a photon's
direction of approach. If we consider clouds
with the same total number of particles, this allows us to to write
\be
A(\vr) \propto V^{\frac{2}{3}}(\vr)
\propto \frac{1}{n_{p,c}^{\frac{2}{3}}(\vr)} ,
\ee
where $n_{p,c}$ is the number density of particles in the cloud.

Thermal equilibrium gives a relation between the number density of
particles in the cloud and the number density of particles in the disc, $n$,
\be
n_{p,c}(\vr) = \frac{n(\vr) T(\vr)}{T_c(\vr)} ,
\ee
where $T$ and $T_c$ are the temperatures of protons in the disc
and the clouds, respectively.
We take the temperature of the clouds to be the same everywhere -
around $T_c = 10^5$K (Kuncic et al, 1996), although the actual
value will be absorbed into an overall constant - and find the temperature of the disc
from the sound speed, $c_s$, as
\be
T(\vr) = \frac{m_p c_s^2(\vr)}{k} ,
\ee
where $k$ is Boltzmann's constant and $m_p$ is the mass of the proton.
This gives us the cross section
\be
A(\vr) \propto
\left[ \frac{k T_c}{m_p n(\vr) c_s^2(\vr)} \right]^{\frac{2}{3}}
\propto \left[ \frac{1}{\rho(\vr) c_s^2(\vr)} \right]^{\frac{2}{3}}.
\ee
We take $\rho(\vr)$ and $c_s^2(\vr)$ from the solutions of
the previous section.
The expression should be normalised by the cross section of the clouds being
around $10^{10}$ cm$^2$ at the innermost radius (Kuncic et al., 1996),
however in practice we can absorb this constant into the constant $\Gamma$,
defined below.

The various assumptions above about the cloud structure
(constant number of particles, constant temperature) result in clouds
with size growing strongly with $r$, as $r^{\frac{5}{3}}$. Combined with
the radial dependence of the flux emission (discussed in more detail below),
this results in profiles dominated by the outer disc regions.
We also want to consider more general radial dependencies,
where profiles are dominated by the central regions.
We therefore add the term $r^{-g}$,
to the cloud cross section, where generally $g$ is a small positive number.
This has a somewhat analogous r\'ole to the $r^{-q}$
term in the standard thin disc
models (Fabian et al. 1989), although has a physically different origin.
The final expression is then
\be
A(\vr) \propto
\left[ \frac{1}{\rho(\vr) c_s^2(\vr)} \right]^{\frac{2}{3}} r^{-g}.
\ee

We take the number density of clouds to trace the number density
of protons in the disc, hence
\be
n_c(\vr) = \Gamma \rho(\vr) .
\ee
The constant of proportionality, $\Gamma$, will determine the degree of
shadowing by optically thick clouds. The probability of obscuration, $P(\vr)$,
with maximal value 1, is proportional to $\Gamma$.

\subsection{Emitted flux}

We take the emitted flux to be proportional to the flux
illuminating the clouds, $f_{in}$, the
number density of the clouds, and the cross section of
the clouds.

We take two contributions to the flux illuminating the clouds,
a central illumination (e.g. due primarily to synchrotron radiation in the
innermost regions of the flow for ADAFs Mahadevan 1997) an
X-ray bath throughout the disc (due to Bremsstrahlung radiation).
The central illumination falls of as $\frac{1}{r^2}$ and
the Bremsstrahlung is proportional to the square of the
number density of particles in the disc. So the
emitted flux is given by
\begin{eqnarray}\label{eq:il}
& &f_{out}(\vr,\vc(\vr)) \propto n_c(\vr) A(\vr) f_{in}(\vr) \nonumber \\
& &\propto \rho A(\vr)^{\frac{2}{3}}
\left[ \frac{A \max(-\hat{\vr}\cdot\hat{\vc}(\vr),0)}{r^2}V_p(\vr) +
B \rho^2 \right].
\end{eqnarray}
Here $A$ and $B$ are constants determining the relative importance
of central radiation and the X-ray bath. They will depend on
the mass accretion rate of the system (Mahadevan 1997) and also the geometry of
the clouds. The $\max(-\hat{\vr}\cdot\hat{\vc}(\vr),0)$ term
contains the information that only the side of the cloud
facing the centre of the disc is illuminated by the central flux.
Hence, the corresponding reprocessed flux is only visible for clouds on
the far side of the hole with respect to the observer,
and the amount of flux
seen by the observer decreases as the illuminated side of the clouds
becomes more edge on to the observer. $V_p(\vr)$ is the primary visibility
function, analogous to the visibility function described earlier,
giving the probability of photons emitted from the central
regions reaching the cloud without being obscured by intermediate clouds.
Fortunately, the photon paths from the 
central illumination are radial and so the integration
of equation \ref{eq:int} becomes analytically tractable, using
appropriate expressions for the cross sections and cloud number densities.

To give some visualisation of the processes involved, we
will consider five cross sections of the disc
at different heights above the equatorial plane, as shown in Fig.1.
For each cross section, we compute
a frequency map and a visibility map for both a centrally illuminated
(synchrotron dominated) case and an X-ray bath (Bremsstrahlung dominated)
case. The disc visualised is a self-similar ADAF disc
with $\epsilon=0.1$ and $\alpha = 0.35$, viewed at 60 degrees.
The results are shown in Fig.2, in which the observer is situated to the
right.

\begin{figure}
\epsfig{file=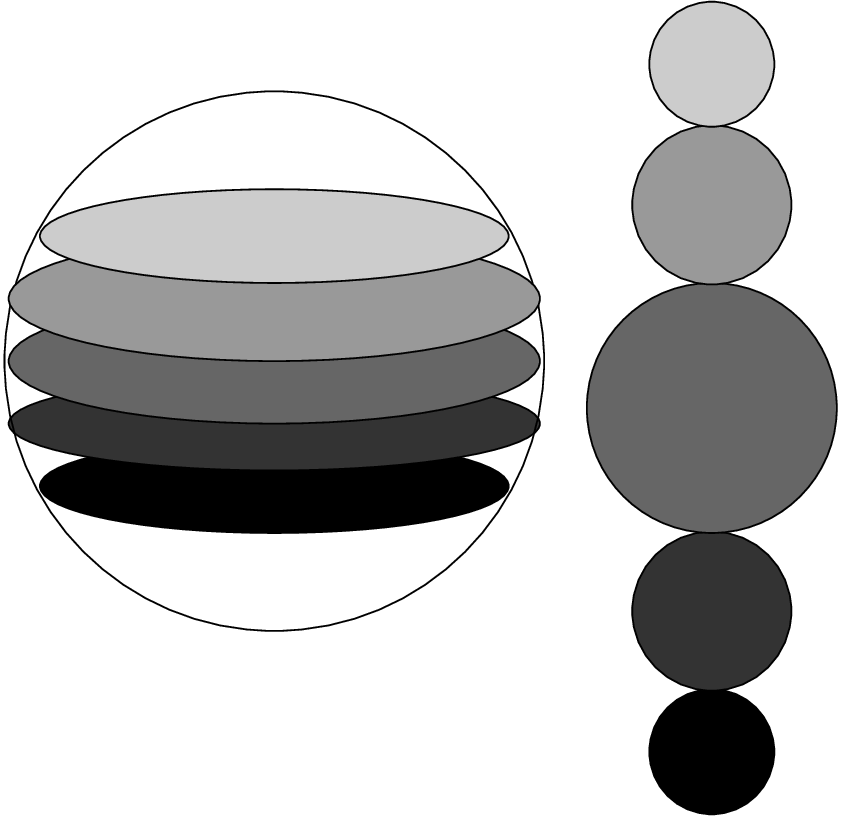,height=2in,angle=270}

\noindent\textbf{Figure 1:} The cross sections of the
disc used in Fig. 2.
\end{figure}

Some thought reveals that the plots of Fig. 2 make sense physically,
and the qualitative features deserve some further comments.
The redshift map of the equatorial is different in two respects
from the standard redshift maps of thin discs. Firstly, the
regions of largest Doppler shift are rotated slightly
with respect to the observer. This is because the radial motion
is of similar magnitude to the rotational motion. Secondly,
gravitational redshift is much more dominant. This is because the
velocities are significantly less than the Keplerian velocities
of a thin disc (see comments in \S 2.1). As we would expect,
the redshifts become less pronounced as we move away from the
equatorial plane.

The visibility in the centrally illuminated case is a superposition
of the visibility in the X-ray bath case with the primary
visibility. The primary visibility implies that the observer
can only see clouds on the far side of the hole (hence there is more
obscuration on the top part of the disc than the bottom part) and
that the more edge-on clouds appear less illuminated.

The visibility in the X-ray bath case is determined
by the size and number density of clouds as well as the
the length of the of the trajectory through the clouds. This
results in the central regions being the most obscured
because the disc under consideration has $g=1.5$, which
means the cloud number density times cross section
is largest in the central regions. Also, the bottom regions
are more obscured than the top regions because they are further
from the observer.

\begin{figure*}

\hspace{0.3in}\epsfig{file=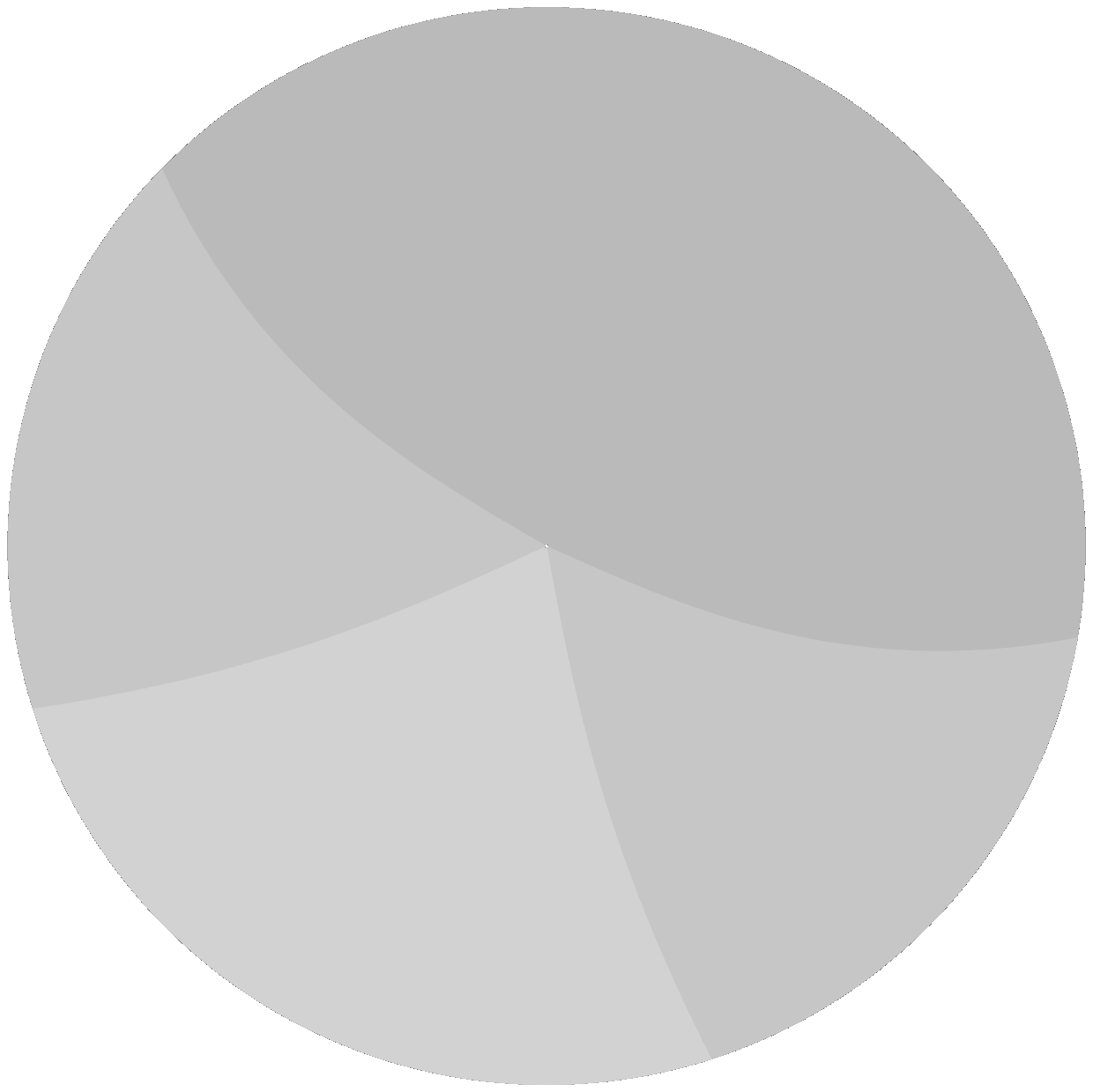,width=1.4in}
\hspace{0.8in}\epsfig{file=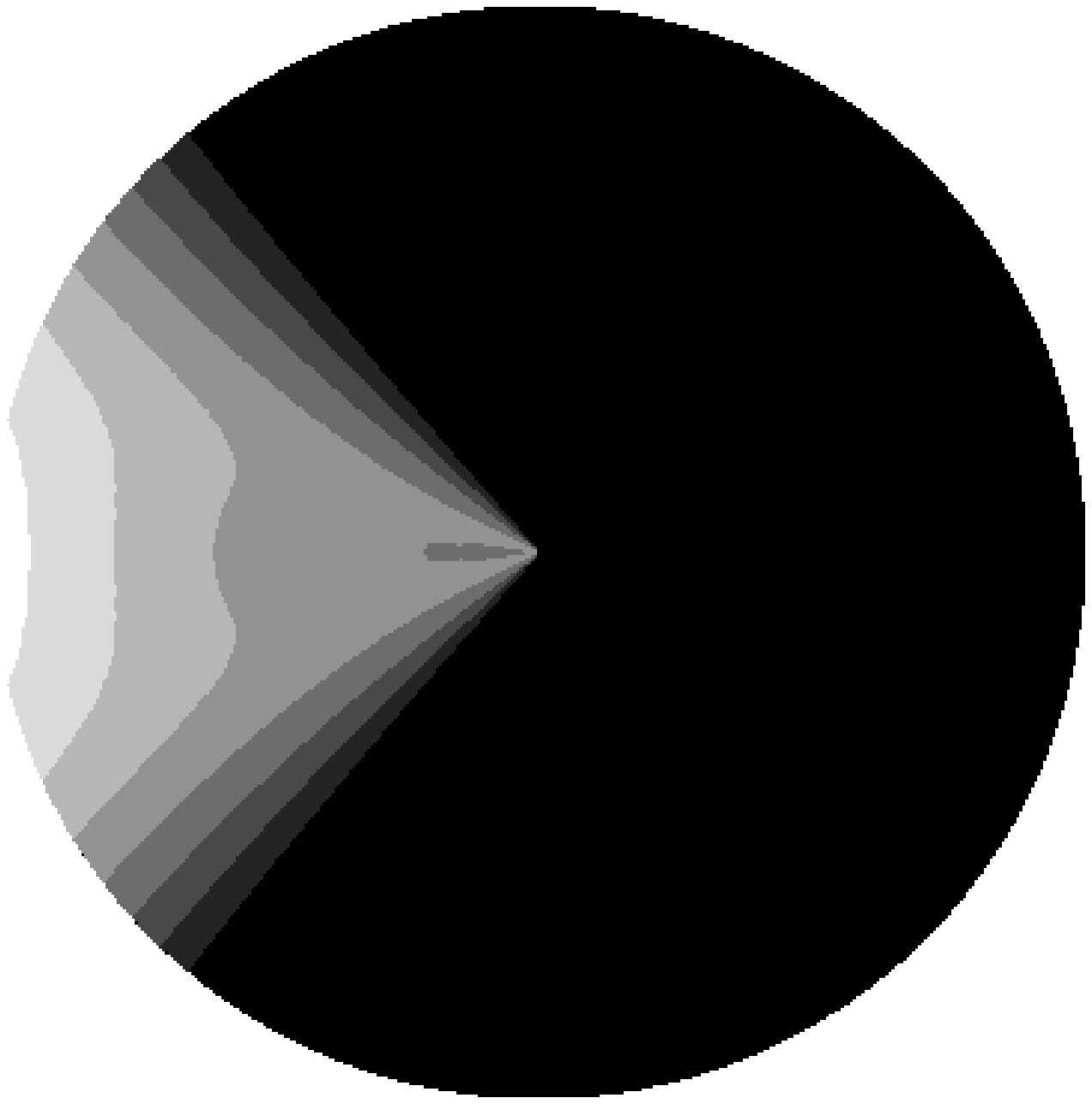,width=1.4in}
\hspace{0.8in}\epsfig{file=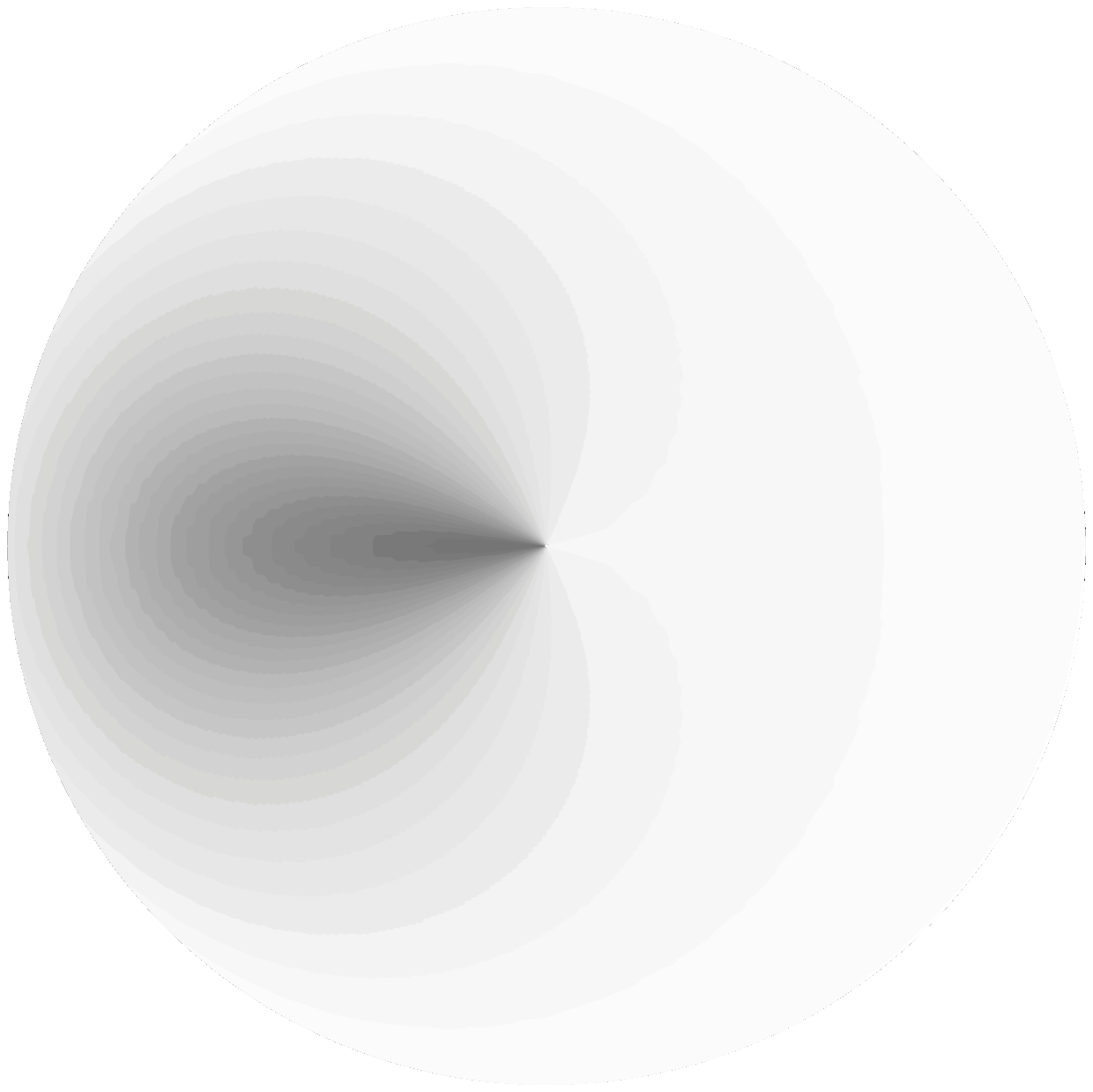,height=1.4in}\hspace{0.3in}

\hspace{0.03in}\epsfig{file=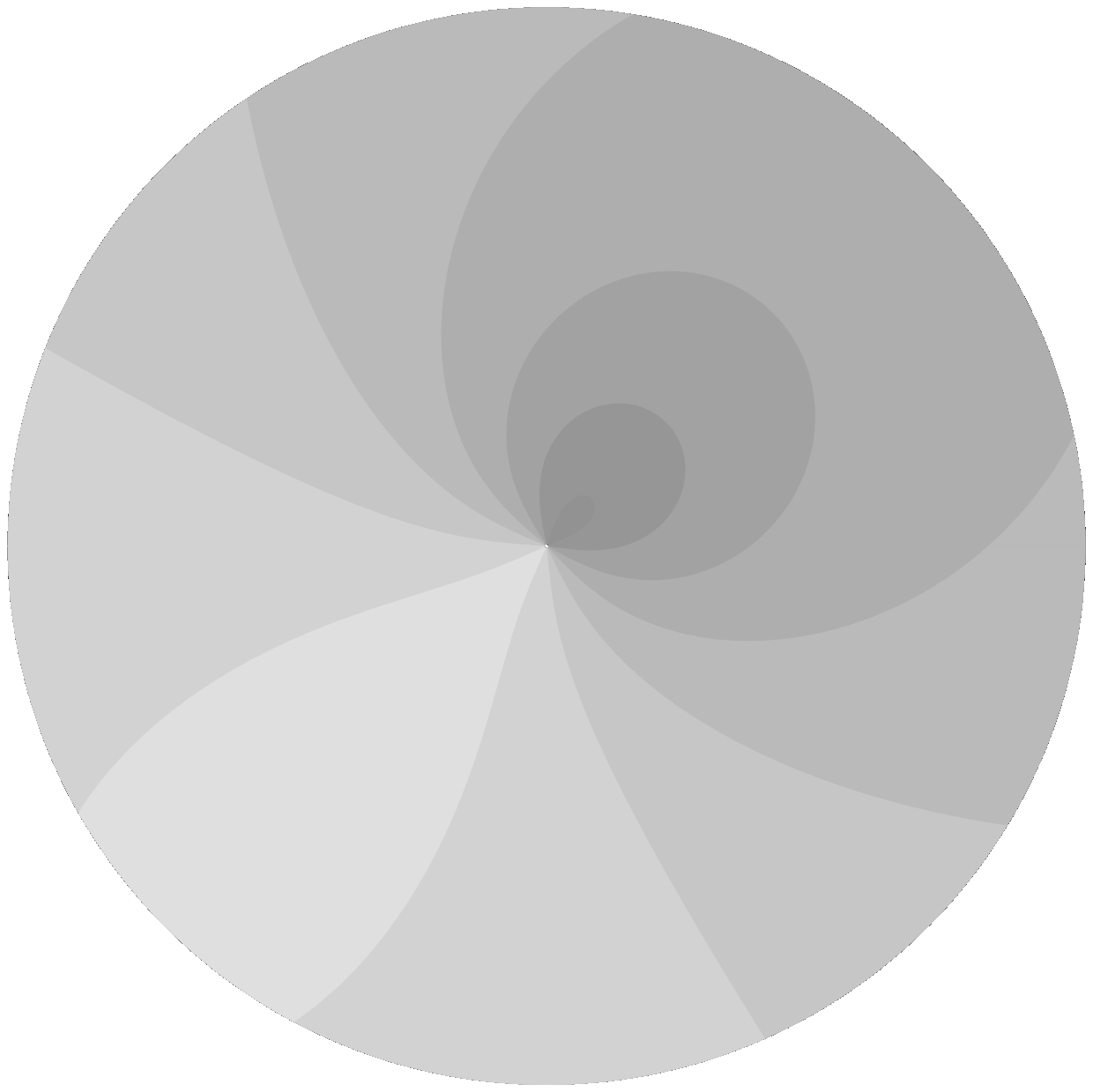,width=1.92in}
\hspace{0.26in}\epsfig{file=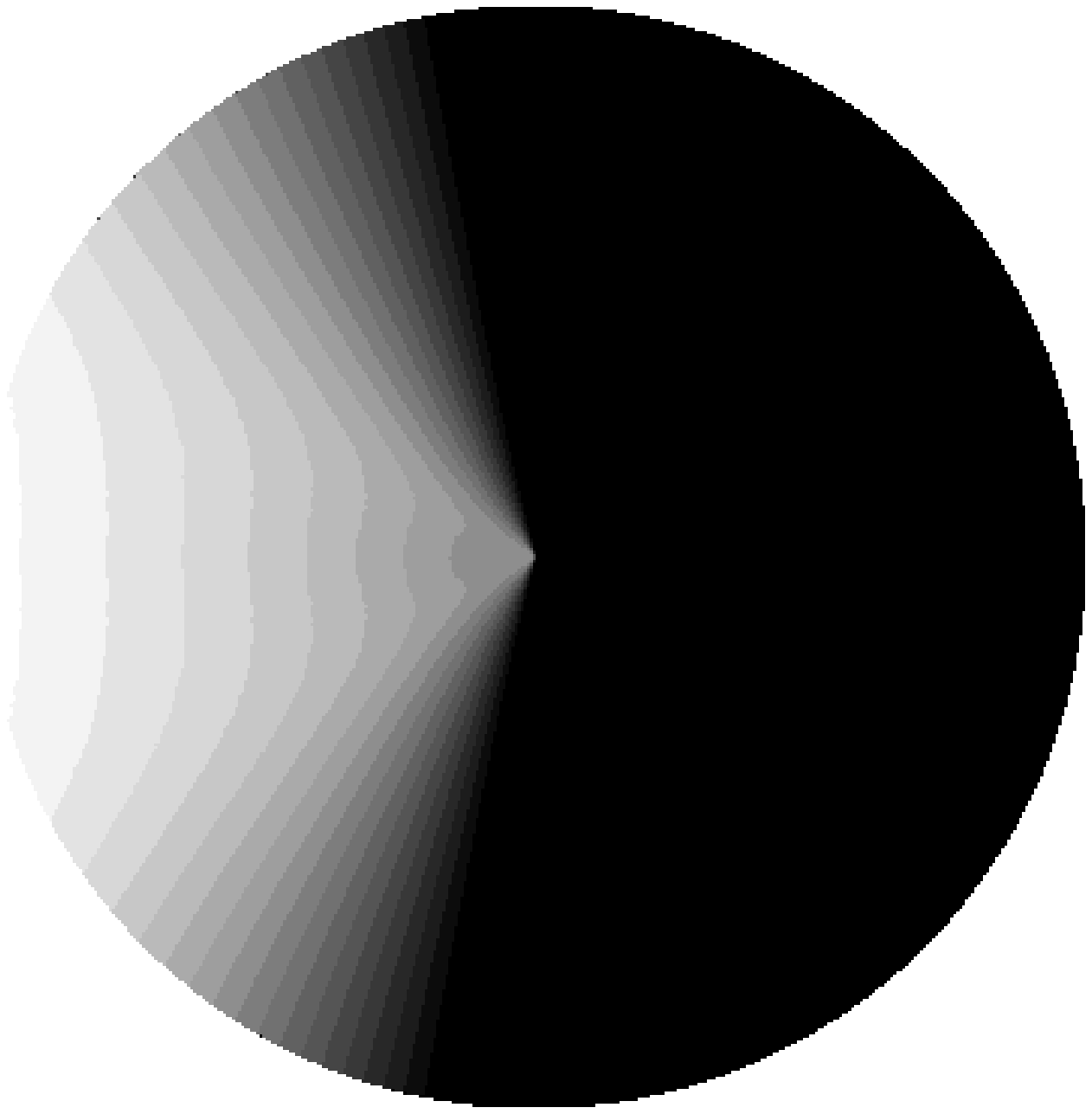,width=1.92in}
\hspace{0.26in}\epsfig{file=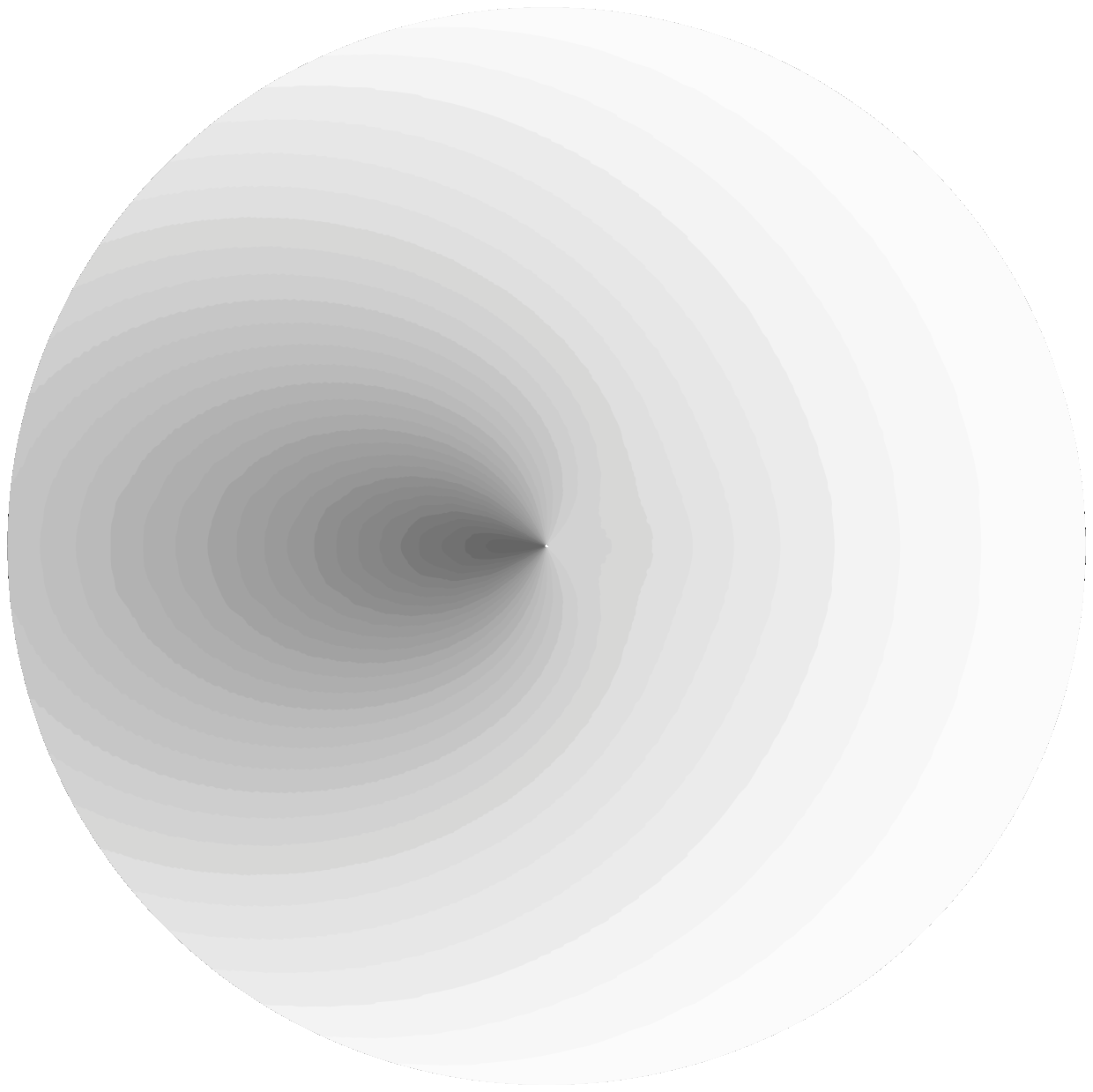,height=1.92in}\hspace{0.03in}

\epsfig{file=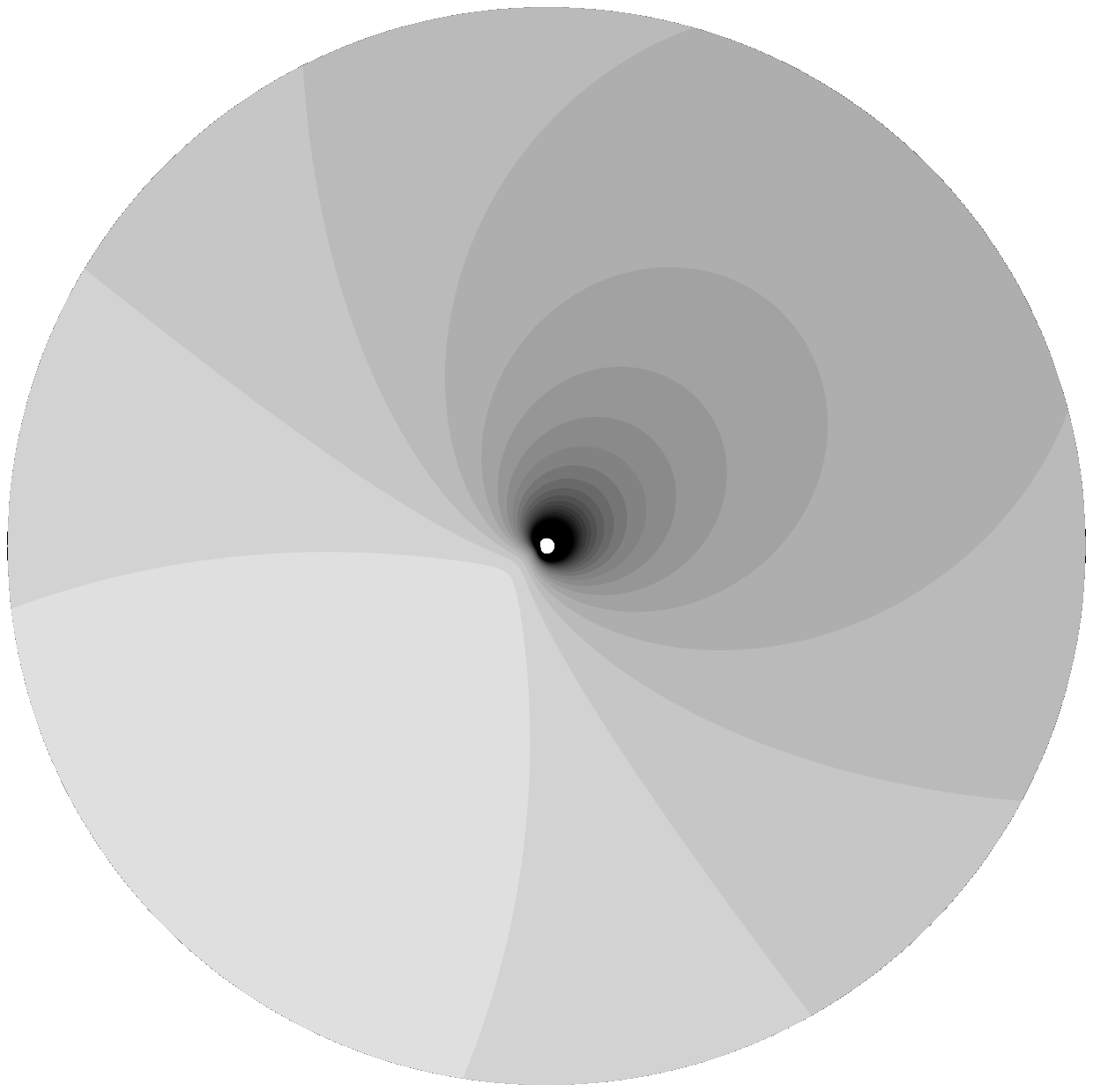,width=2in}\hspace{0.2in}
\epsfig{file=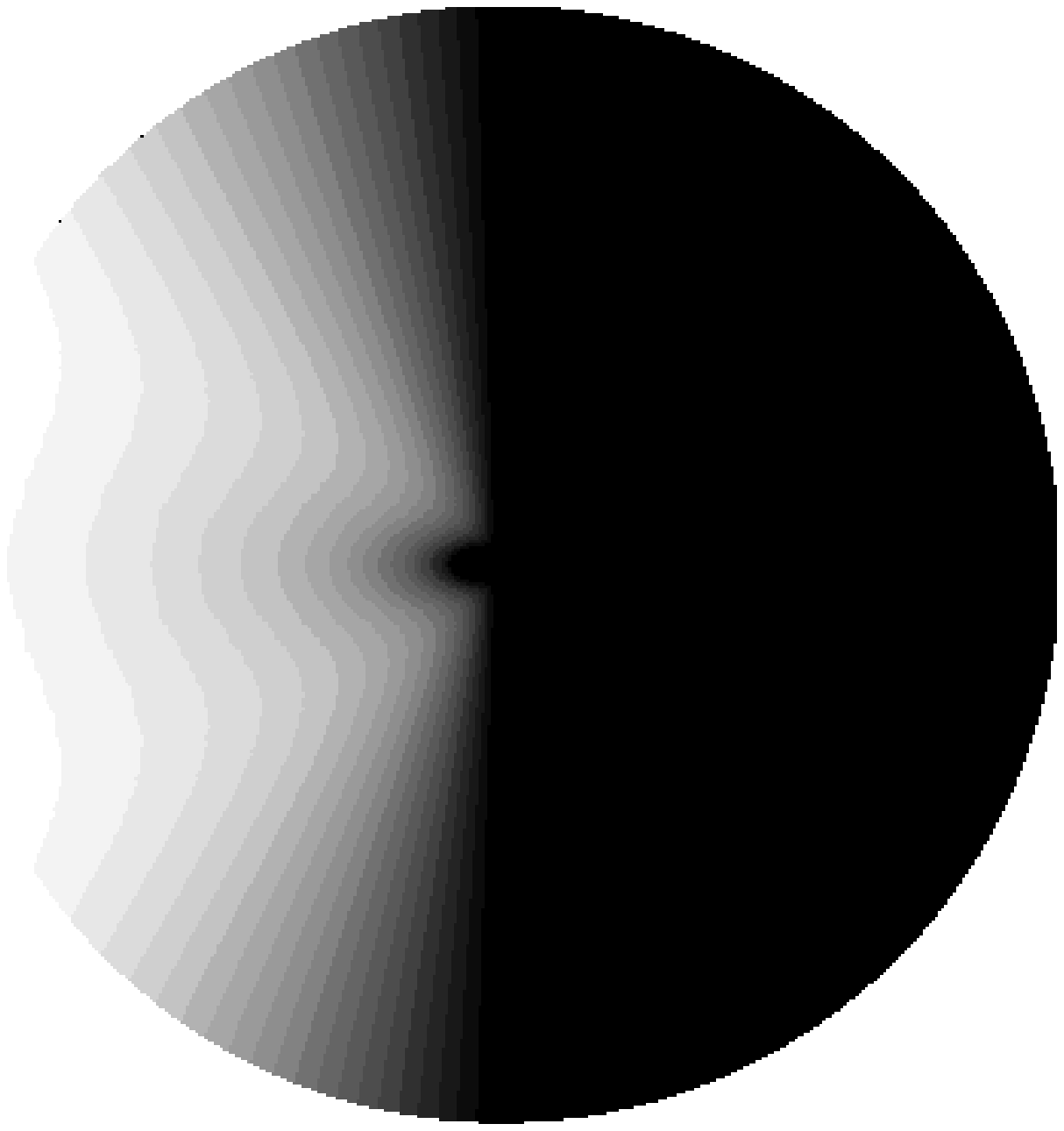,width=2in}\hspace{0.2in}
\epsfig{file=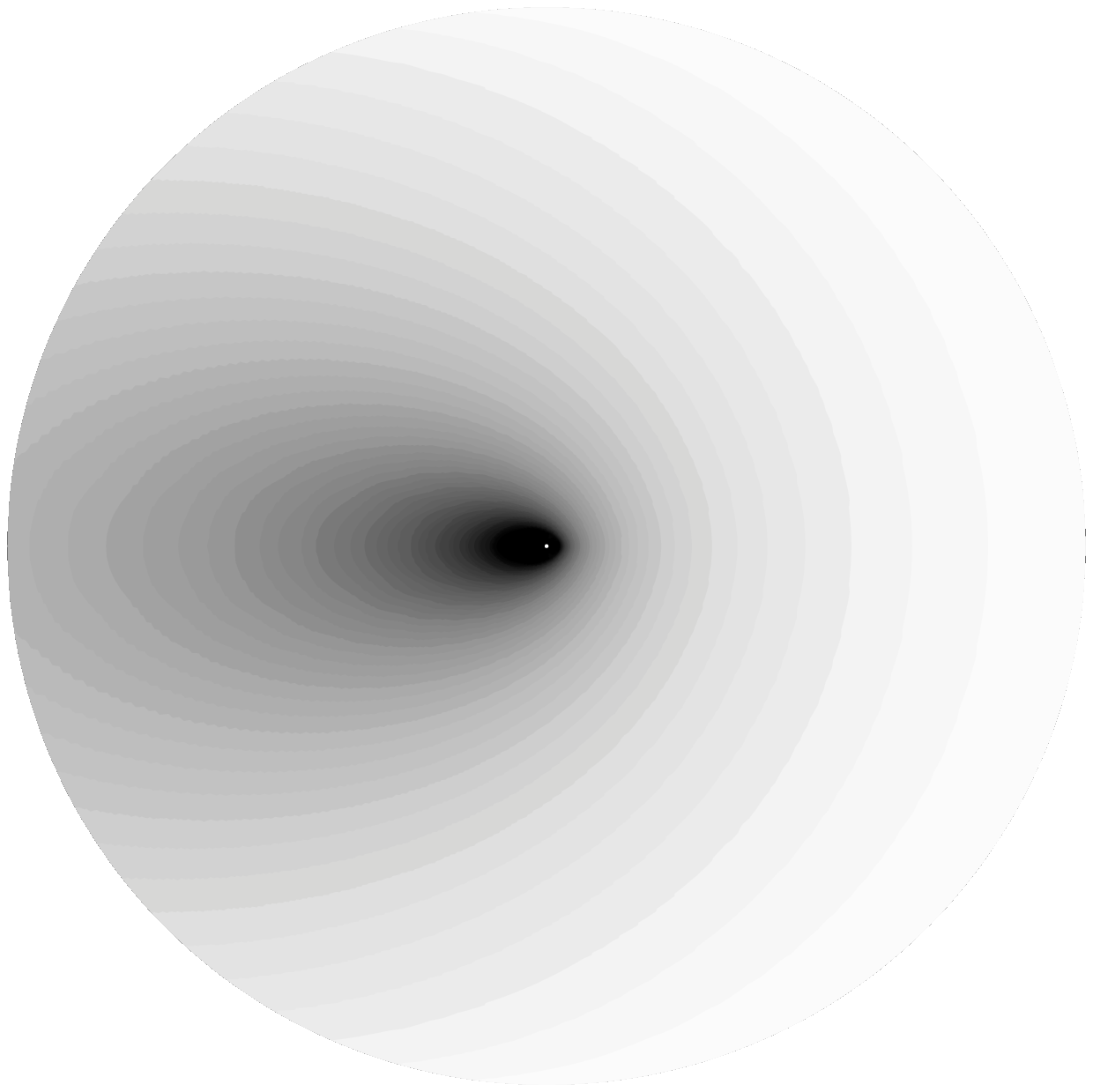,width=2in}

\hspace{0.03in}\epsfig{file=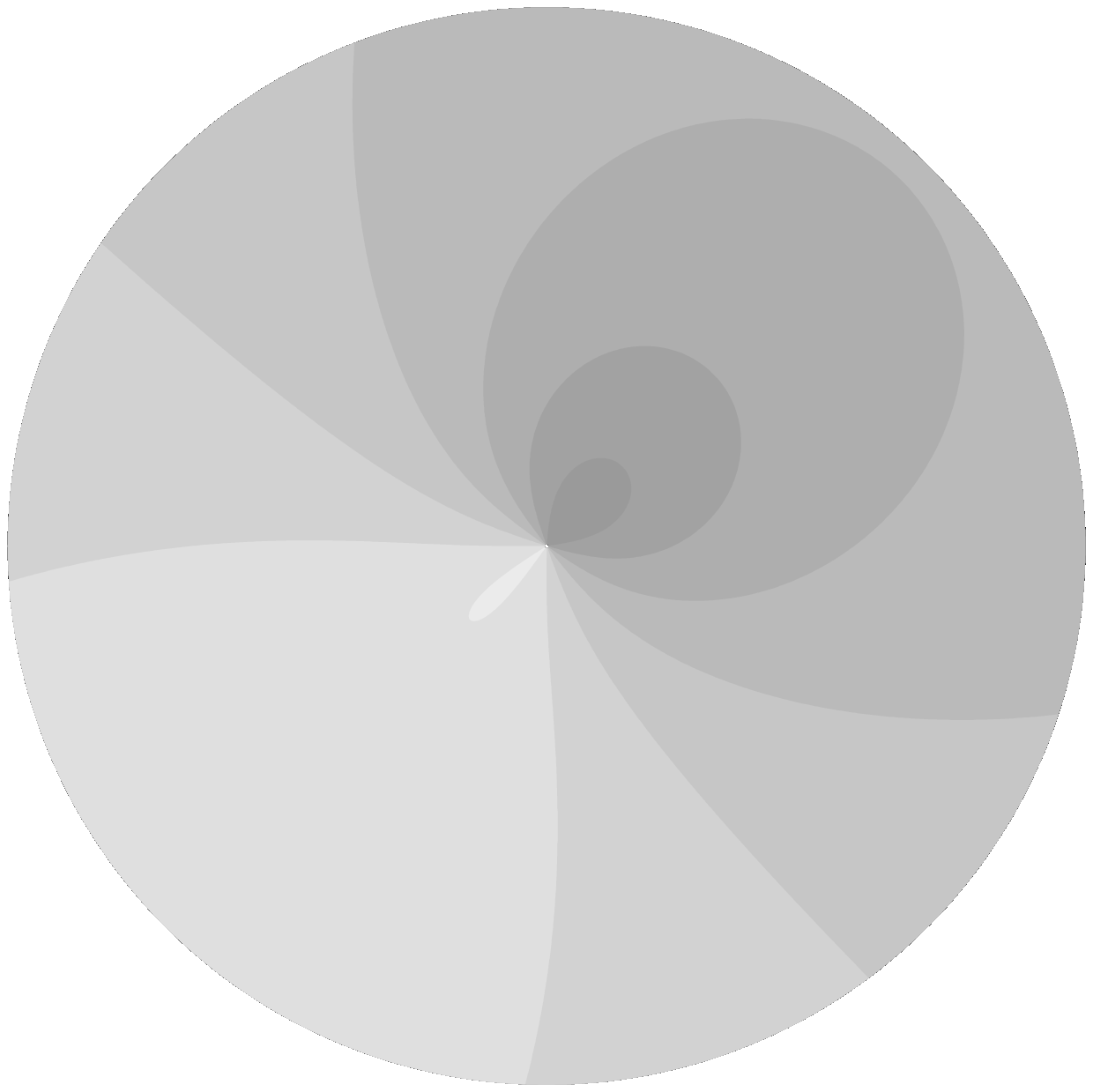,width=1.92in}
\hspace{0.26in}\epsfig{file=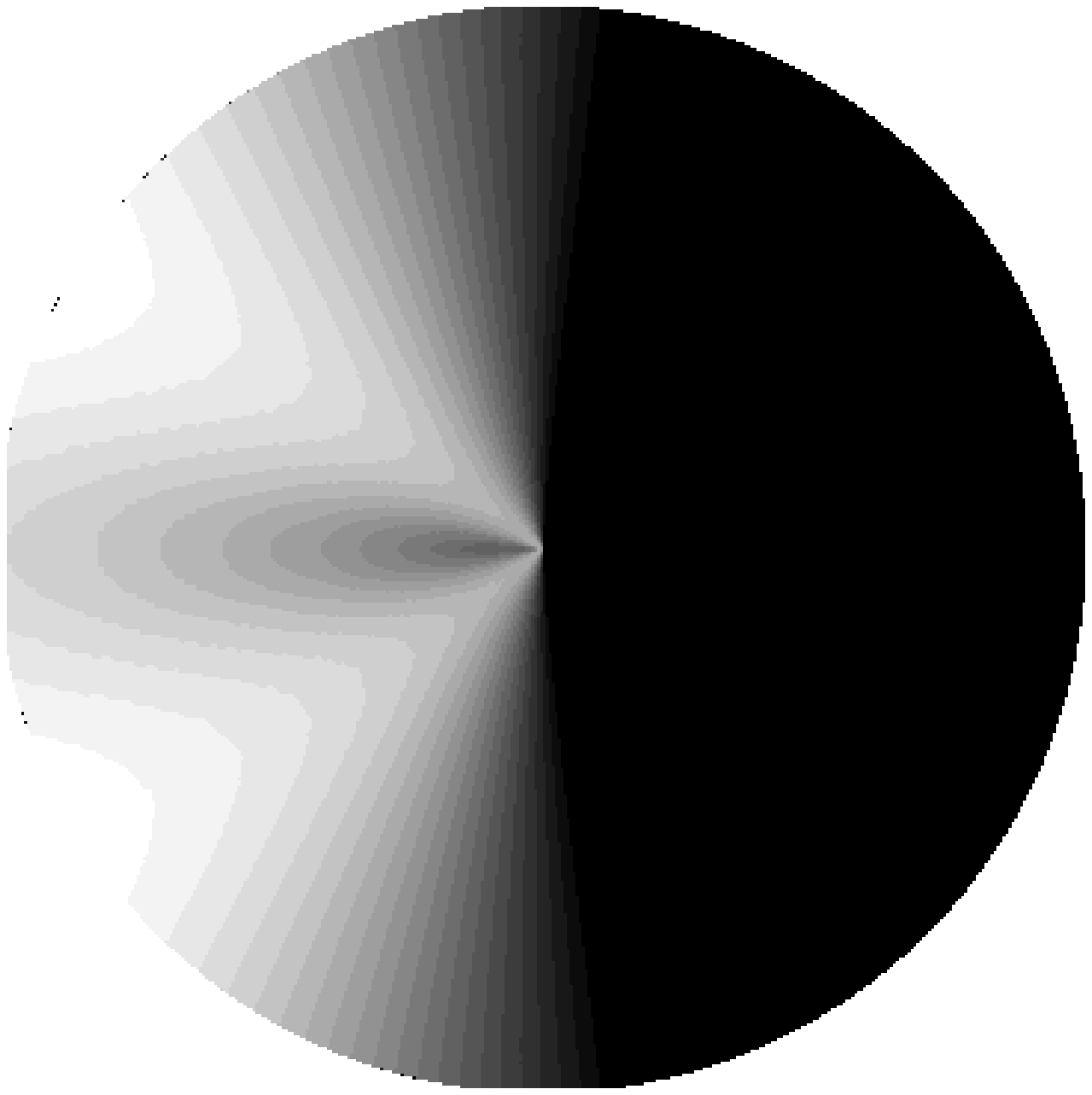,width=1.92in}
\hspace{0.26in}\epsfig{file=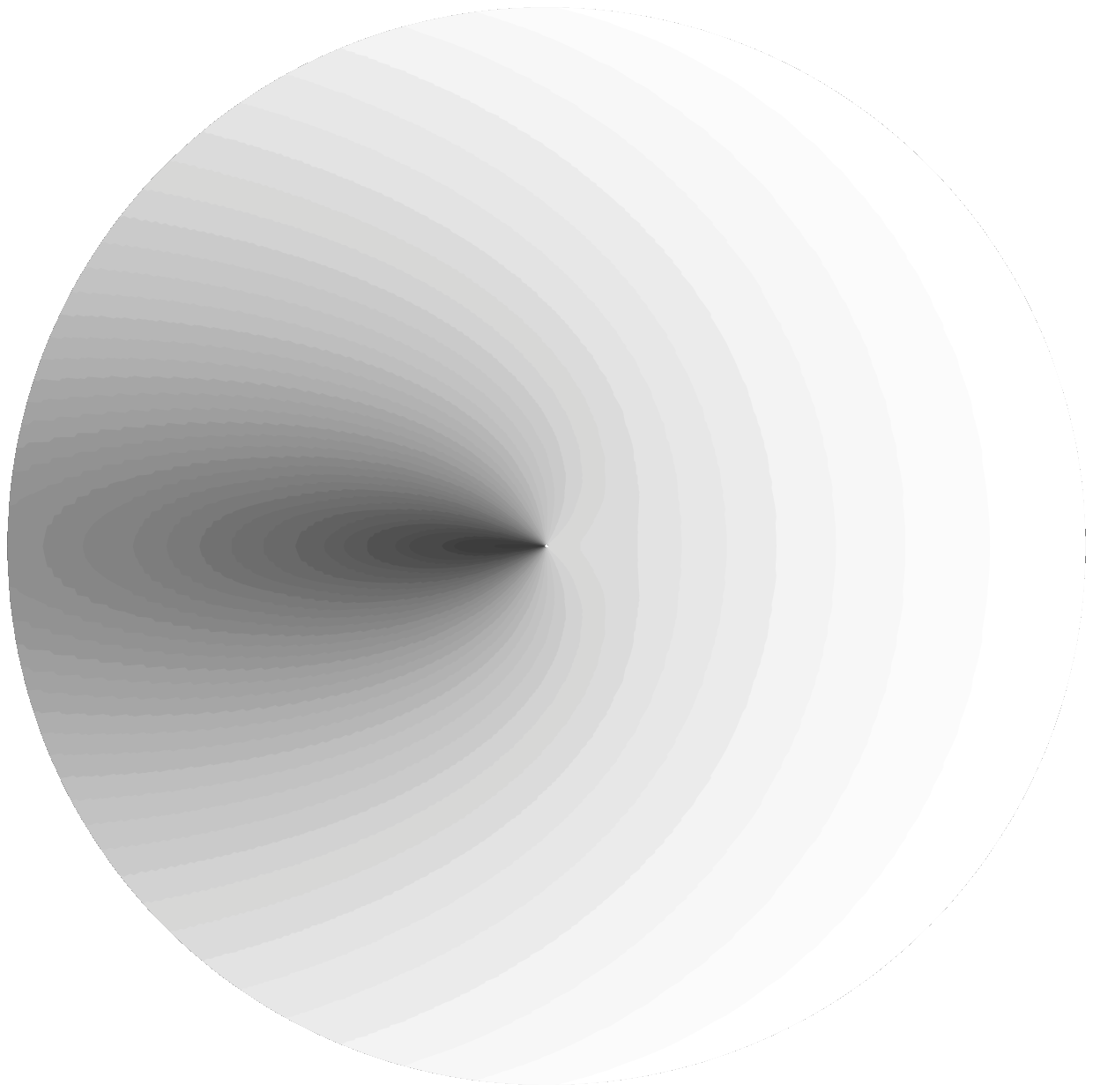,width=1.92in}\hspace{0.03in}

\hspace{0.3in}\epsfig{file=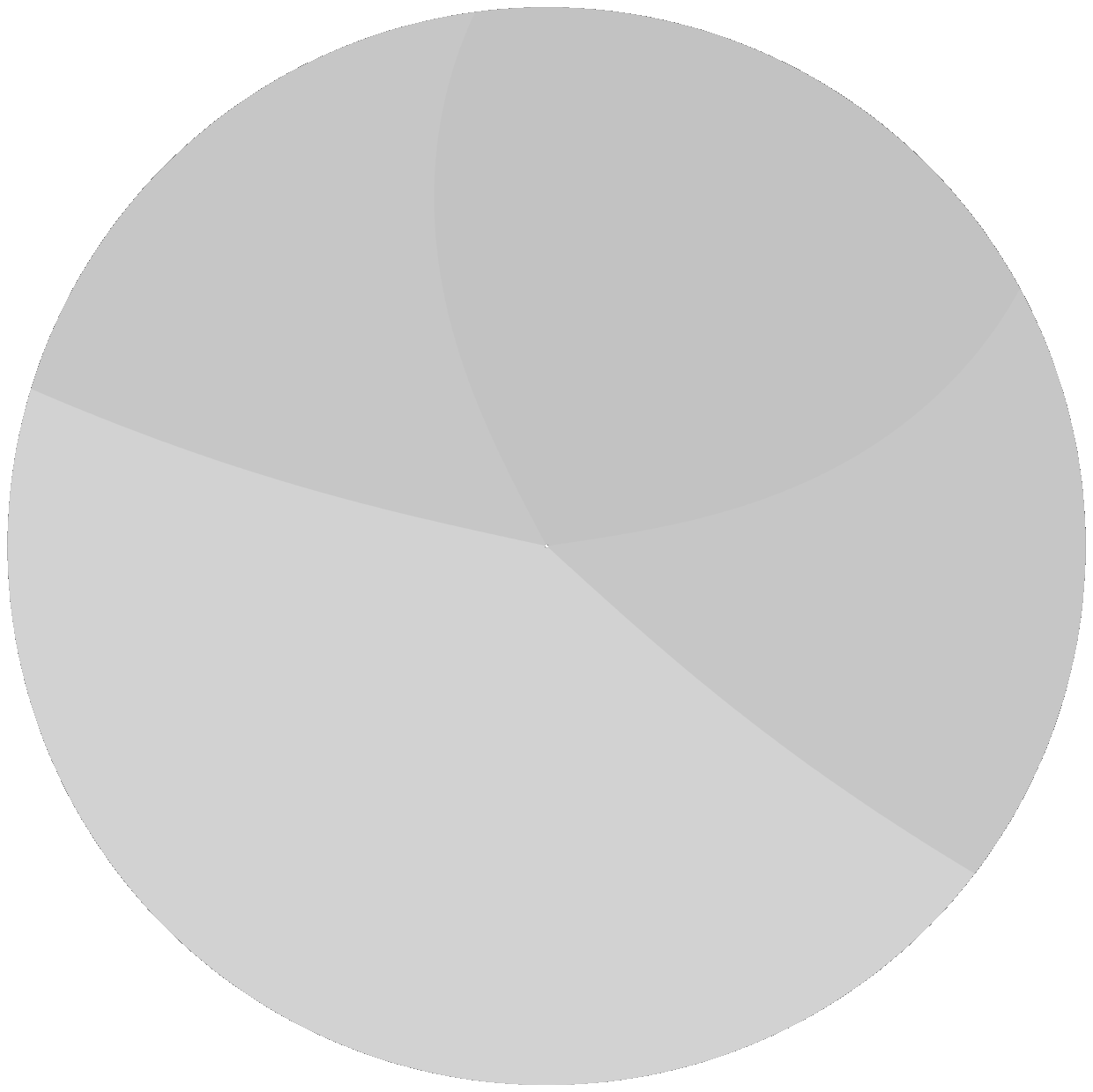,width=1.4in}
\hspace{0.8in}\epsfig{file=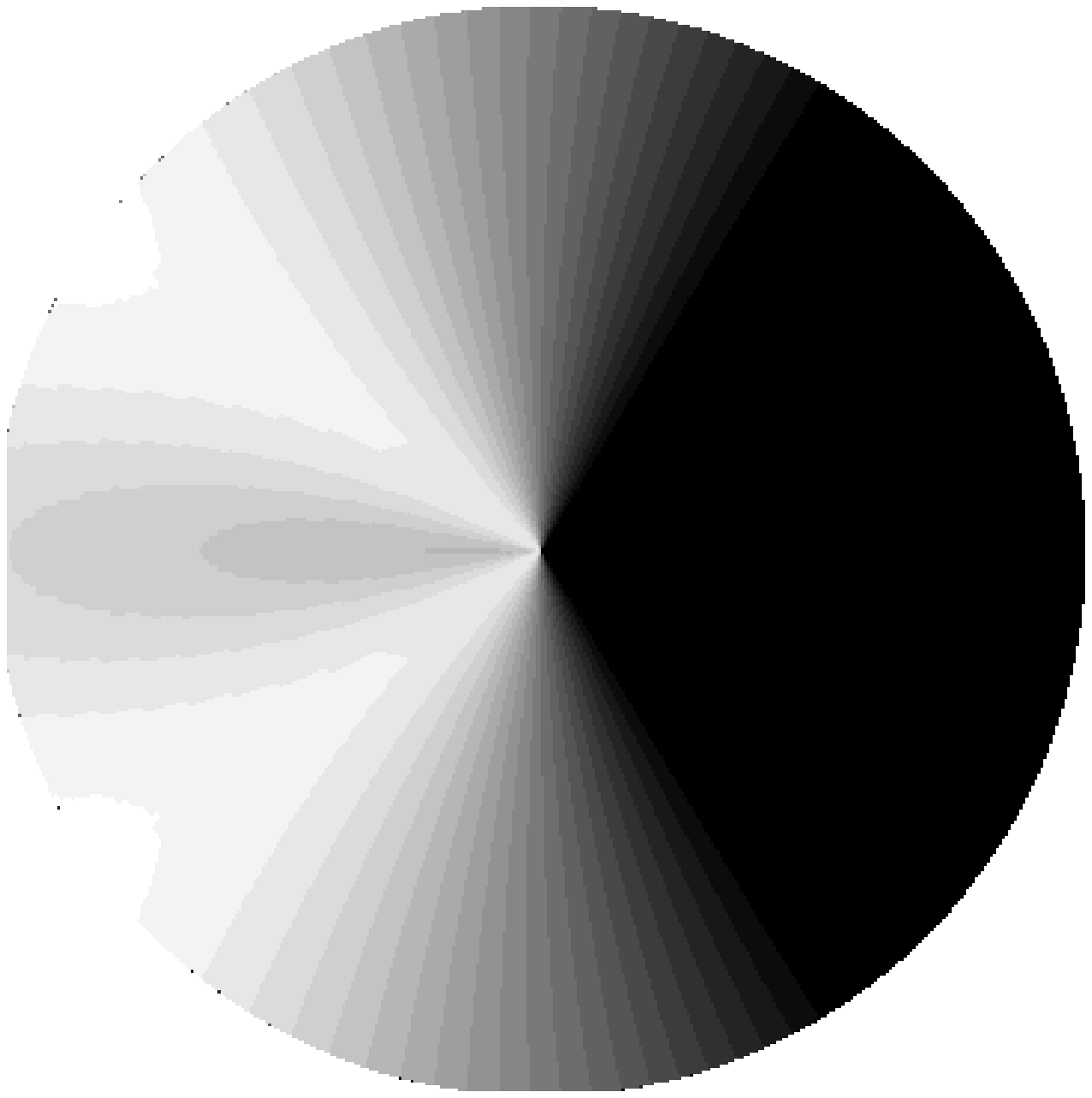,width=1.4in}
\hspace{0.8in}\epsfig{file=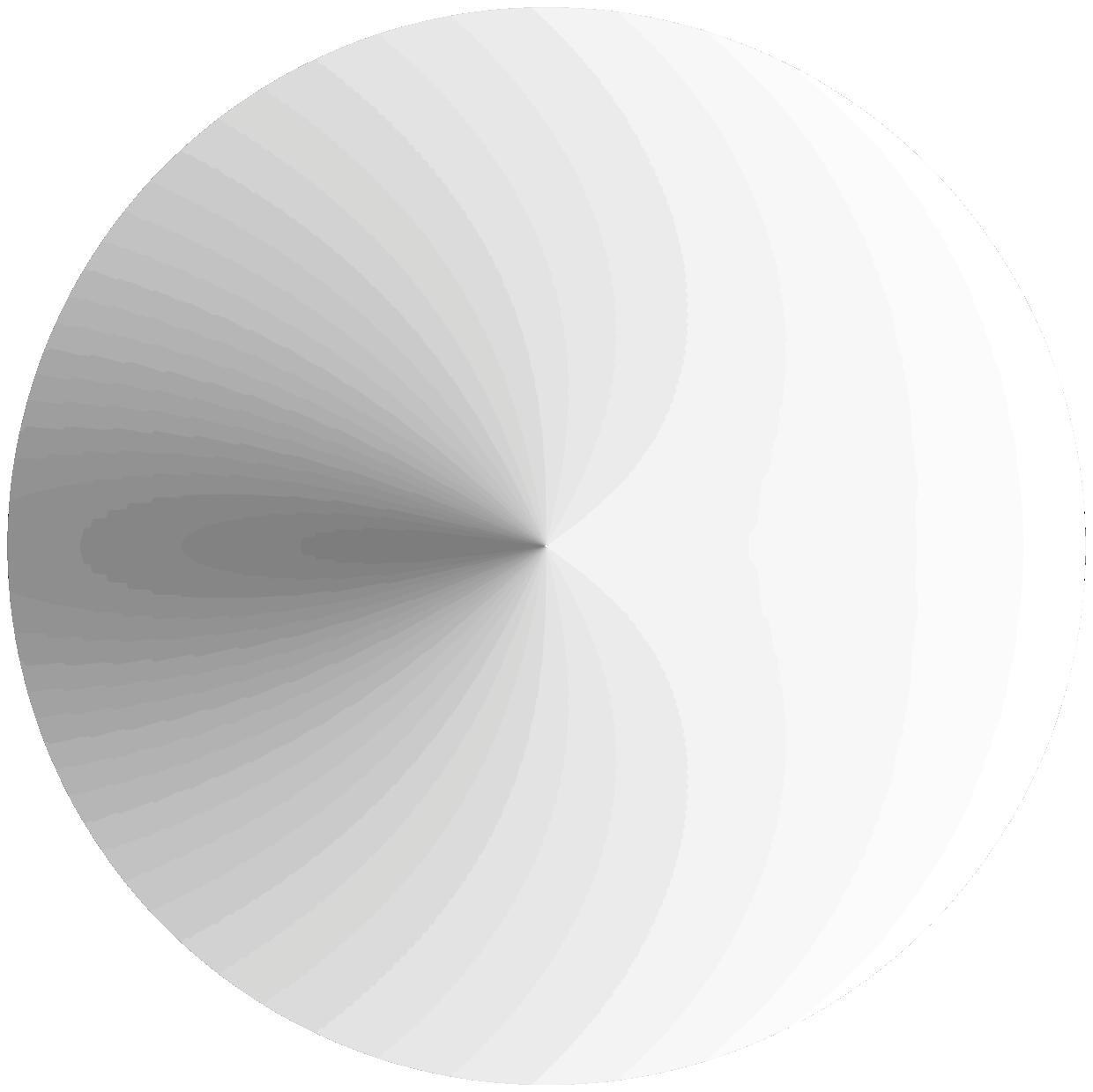,width=1.4in}\hspace{0.3in}

\noindent\textbf{Figure 2:} From top to bottom,
cross sections of a disc at heights of
$0.7r_{out}$, $0.25r_{out}$, $0$, $-0.25r_{out}$ and $-0.7r_{out}$
vertically relative to the equatorial plane, drawn to scale. The left column shows
redshifts (darker shows more redshifted),
the central column shows visibility in the centrally illuminated
(synchrotron dominated) case and the right column shows
visibility in an X-ray bath (Bremsstrahlung dominated) case. White
regions are totally unshadowed and black regions are totally shadowed.
The observer is to the right and the disc is rotating counter clockwise.
\end{figure*}

\section{Line profiles}

Discs were considered $(i)$ at two or three different
inclinations (15,30 and 60 degrees, or some subset thereof),
$(ii)$ once with shadowing ($\Gamma$ finite)
and once without ($\Gamma=0$), and $(iii)$
once with the profile dominated by the outer regions ($g=0$)
and once with the profile dominated by inner regions ($g=1.5$).
All discs have $\alpha = 0.35$.

\subsection{Self-similar ADAF profiles}

We consider three ADAF discs. The first 
has $\epsilon=0.1$, hence the density dependence is essentially spherically
symmetric. We take this disc to be
dominated by central illumination (i.e. $A=10, B=1$ in
equation \ref{eq:il}). The second disc also has $\epsilon=0.1$ but is
dominated by the X-ray bath (i.e. $A=1,B=100$). The third has
$\epsilon=10$, hence the density is concentrated within around 30 degrees
of $\theta$ on each side of the equatorial plane. We 
take this disc to be dominated by central illumination.

A striking feature of the profiles from the two $\epsilon=0.1$ ADAF
discs considered, Figs. 3 and 4, is the lack of significant dependence
on the inclination angle (compare to the profiles from flat thin discs in,
e.g. Fabian et al. 1989, which show a strong inclination dependence).
This is because the velocities given by equations \ref{eq:two}-\ref{eq:three}
have magnitudes smaller than the Keplerian velocities of thin discs
and so  extreme blue Doppler shifts do not appear. 
There is more
blue-shifted flux for the higher inclination discs,
which is why the 
peak very near the rest frequency
is lower in the higher inclination discs.

The discs with $g=1.5$ have profiles dominated by flux from the inner regions,
which explains the greater presence of gravitationally redshifted flux.
This is most notable in the X-ray bath case of Fig. 4 because here
there is a contribution from Doppler redshifted flux which is not so
visible in the centrally illuminated case. See Fig.2 for how the shadowings
and redshifts interact. The fact that 
$v_r \approx v_{\theta}$ in the discs of Figs. 3 \& 4
is important because it rotates the most Doppler redshifted region into the
less visible sector of the disc for the centrally illuminated case, Fig. 3.

\begin{figure}
\epsfig{file=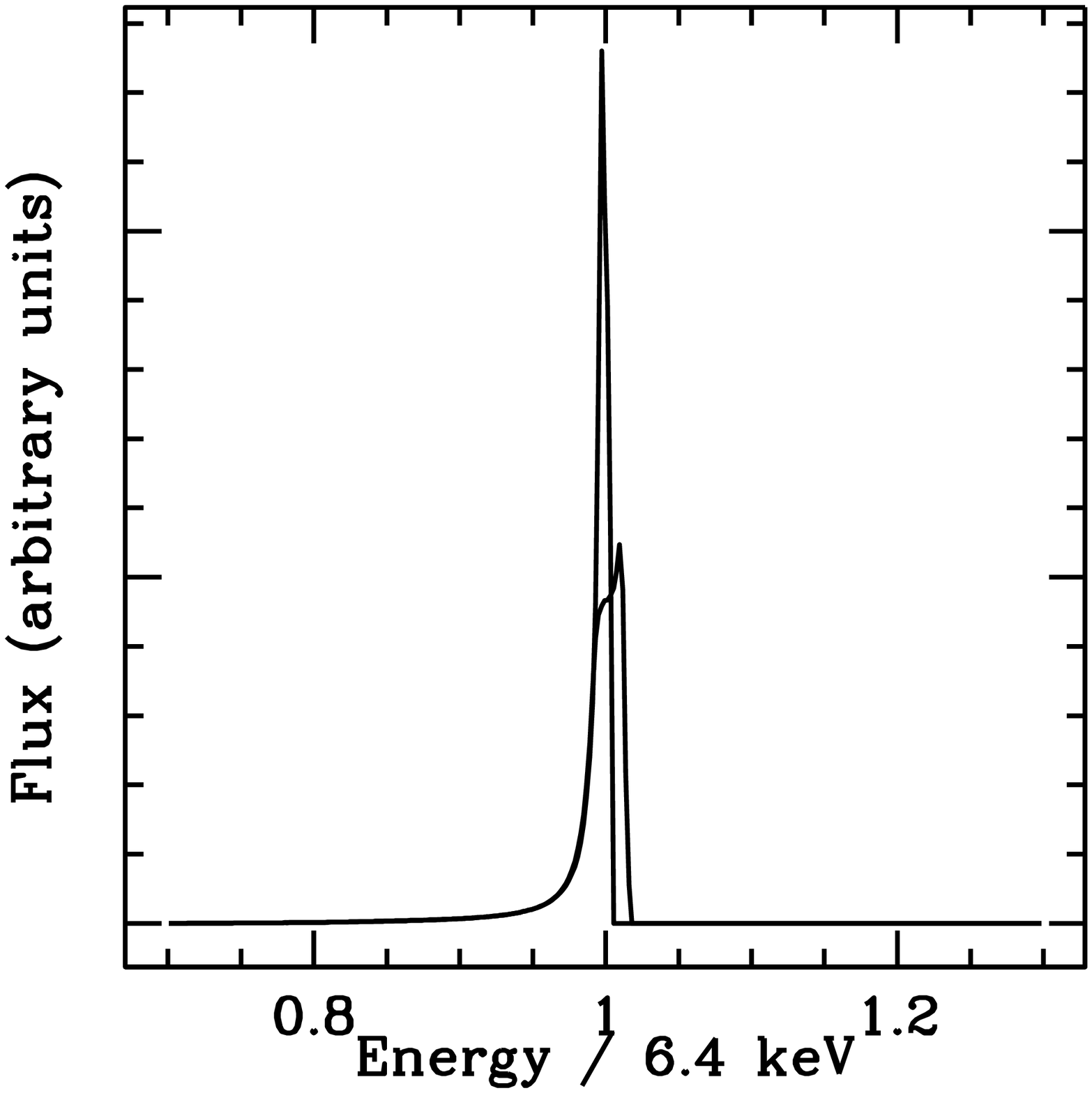,width=1.8in}\epsfig{file=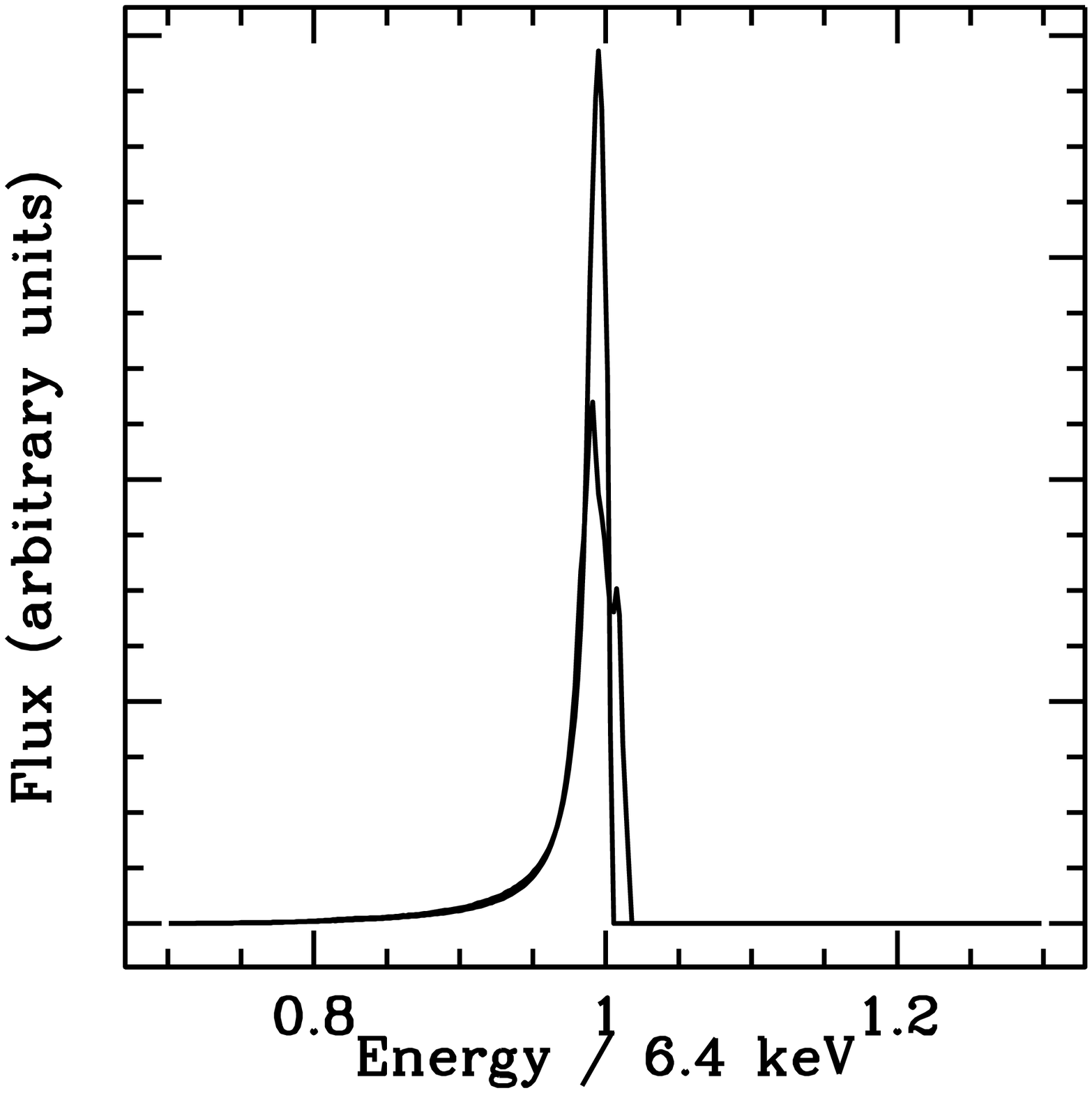,width=1.8in}\\
\epsfig{file=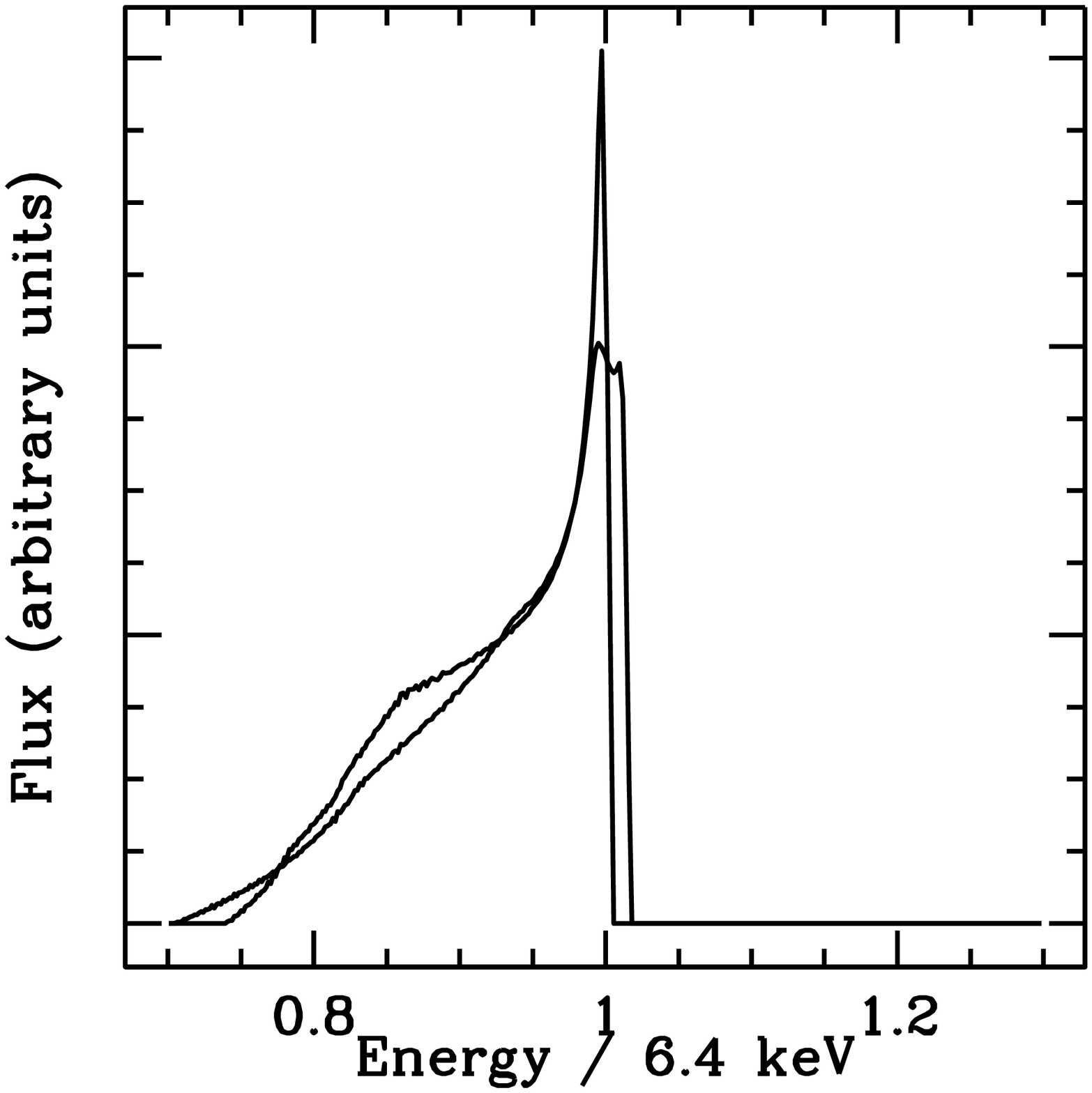,width=1.8in}\epsfig{file=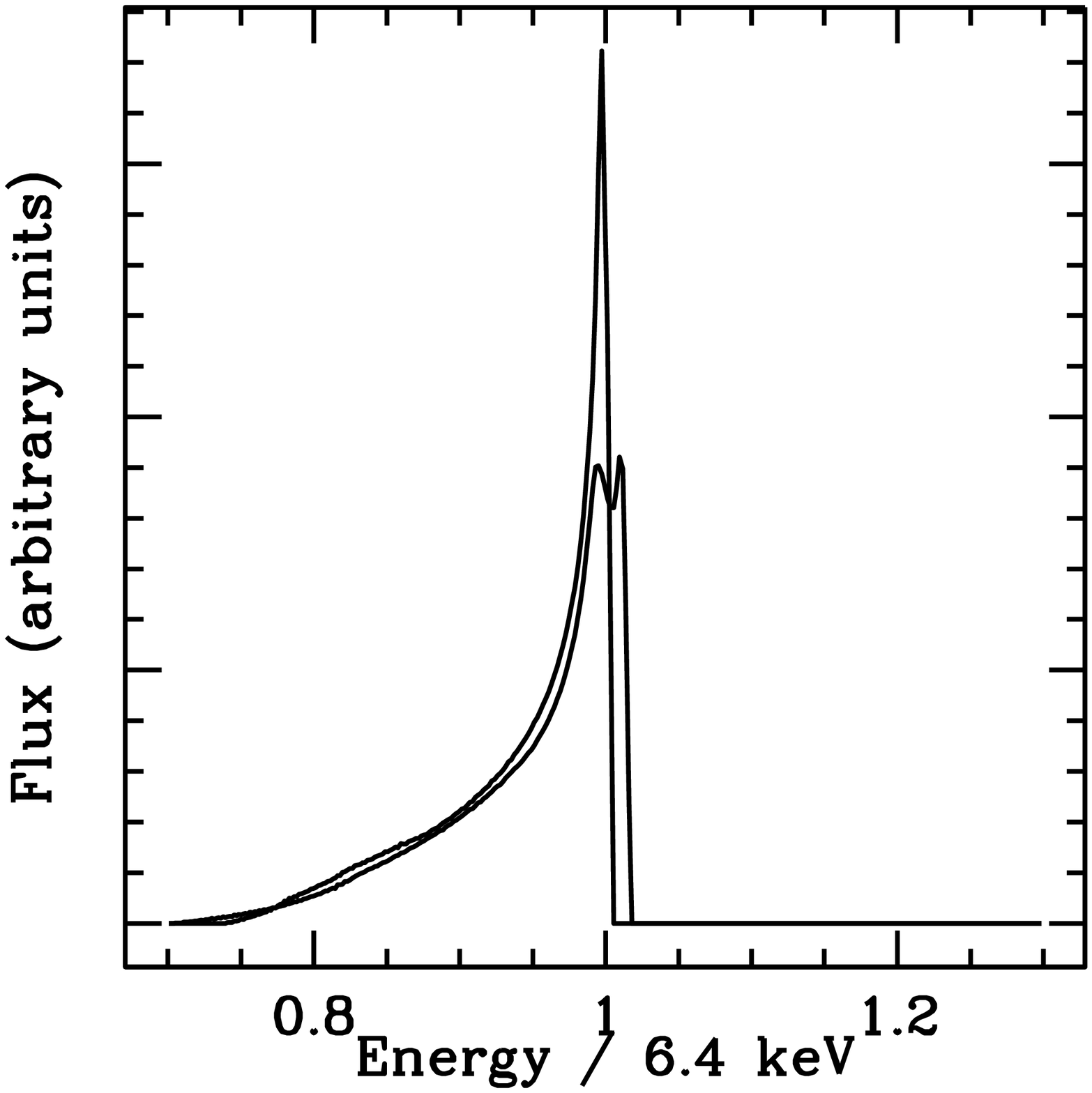,width=1.8in}\\

\noindent\textbf{Figure 3:} Profiles from an ADAF disc with $\epsilon=0.1$.
Central illumination dominated ($A=10$, $B=1$). Left to right:
non-shadowed and shadowed.
Top to bottom: $g=0$ (outer region dominated) and $g=1.5$ (inner region
dominated).
Each plot shows profiles for inclinations of 15 and 60 degrees,
plotted to scale (i.e. not separately normalised). The
profile with the higher peak is at 15 degrees.
\end{figure}

\begin{figure}
\epsfig{file=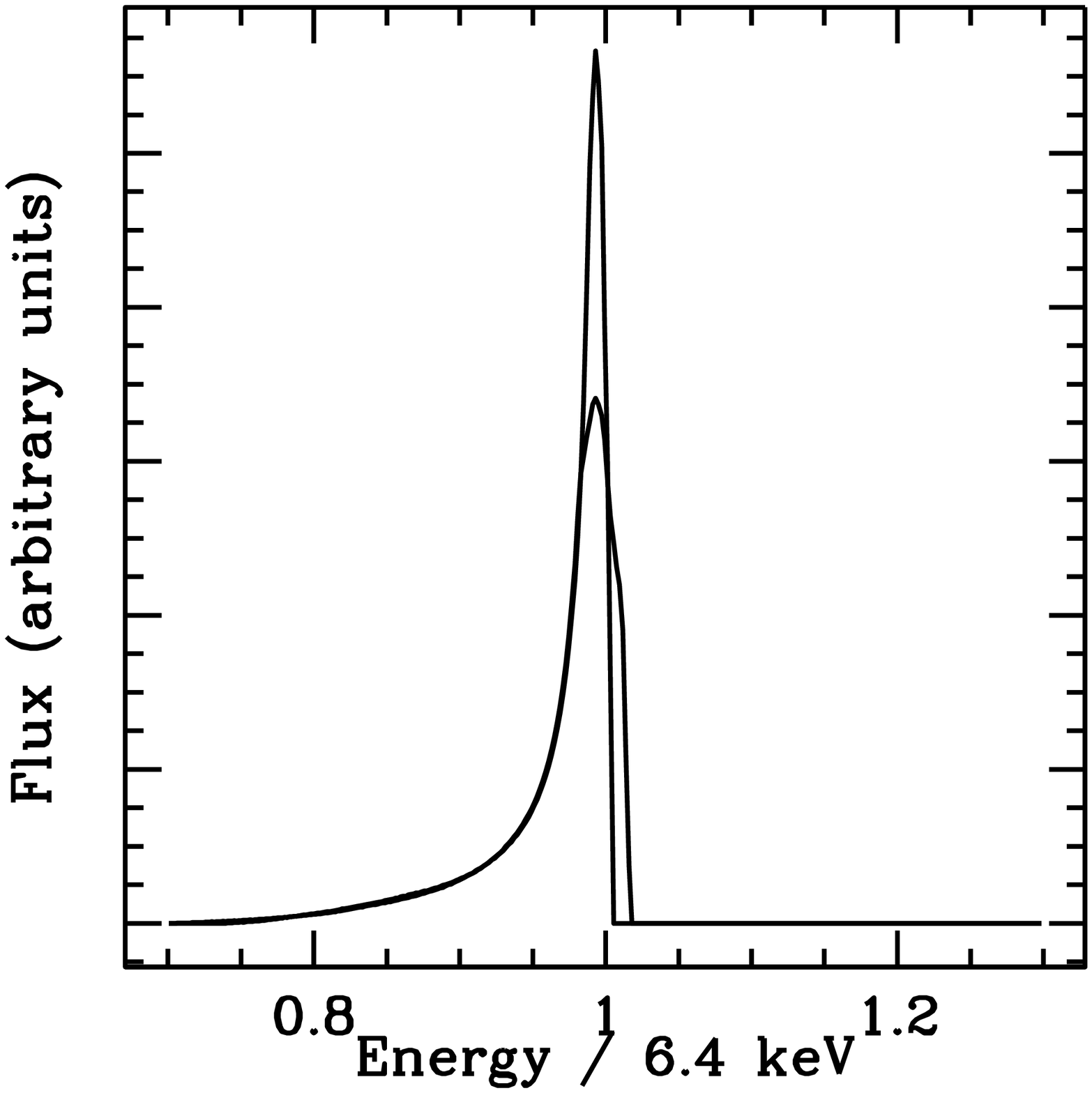,width=1.8in}\epsfig{file=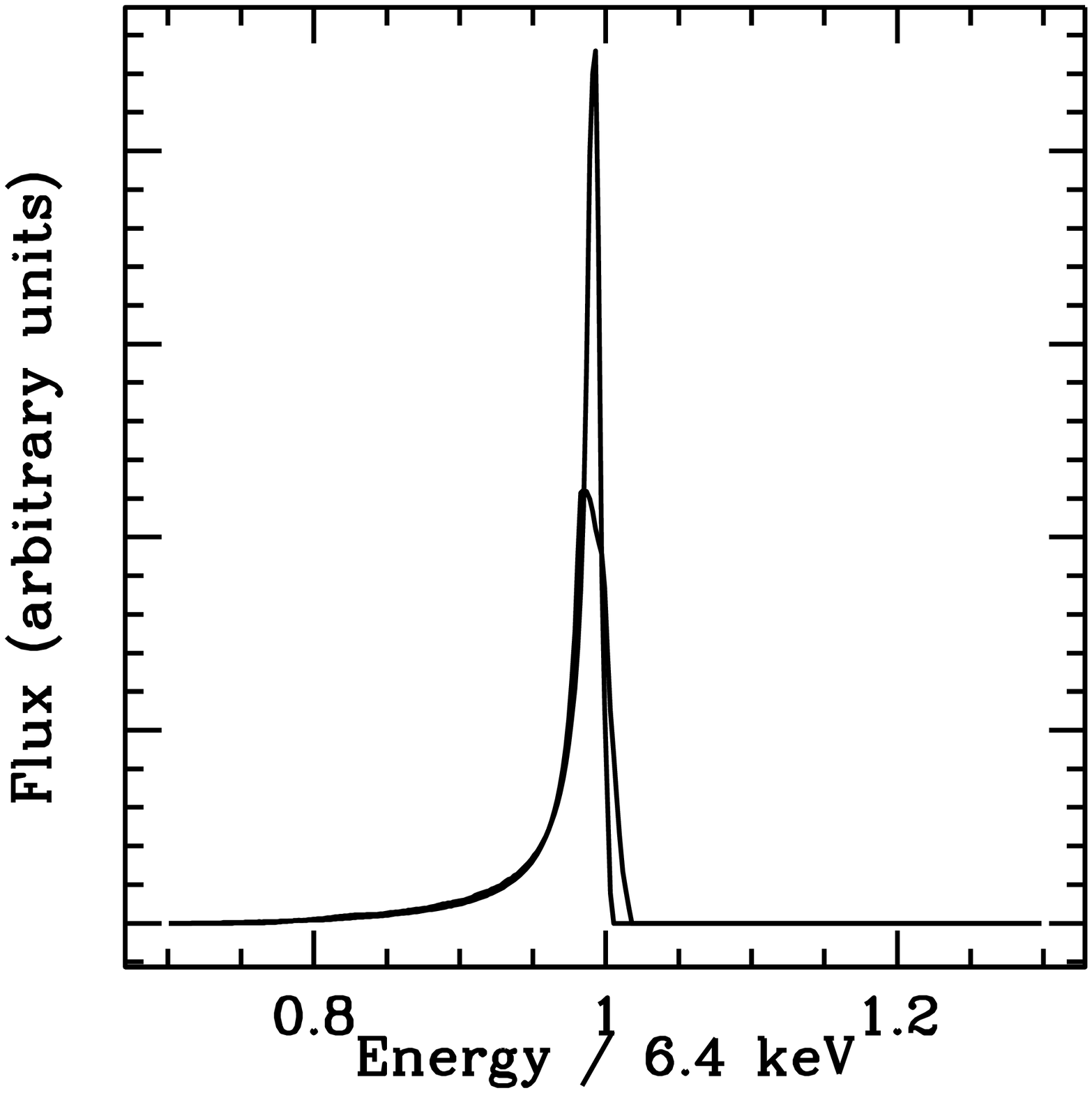,width=1.8in}\\
\epsfig{file=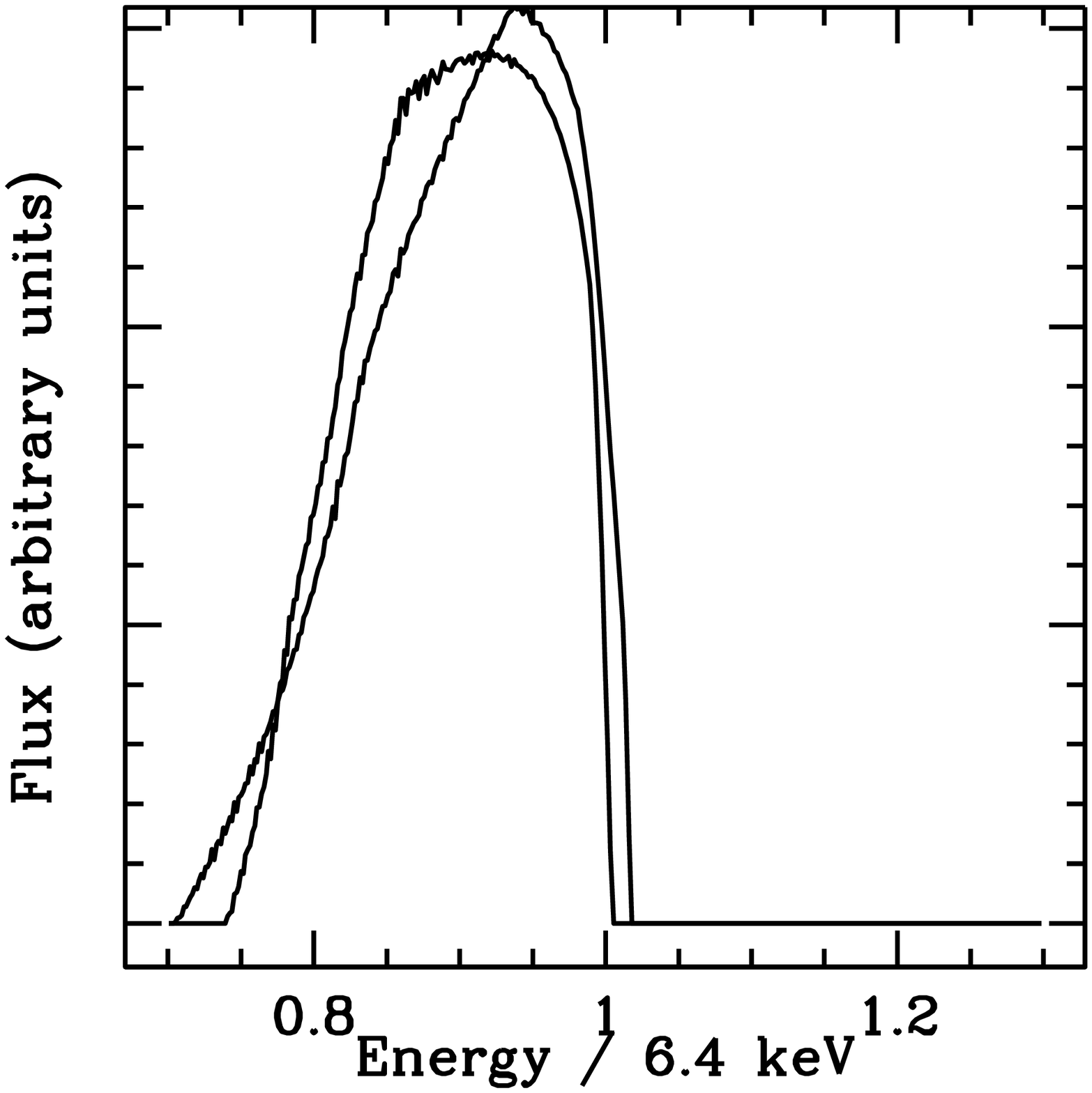,width=1.8in}\epsfig{file=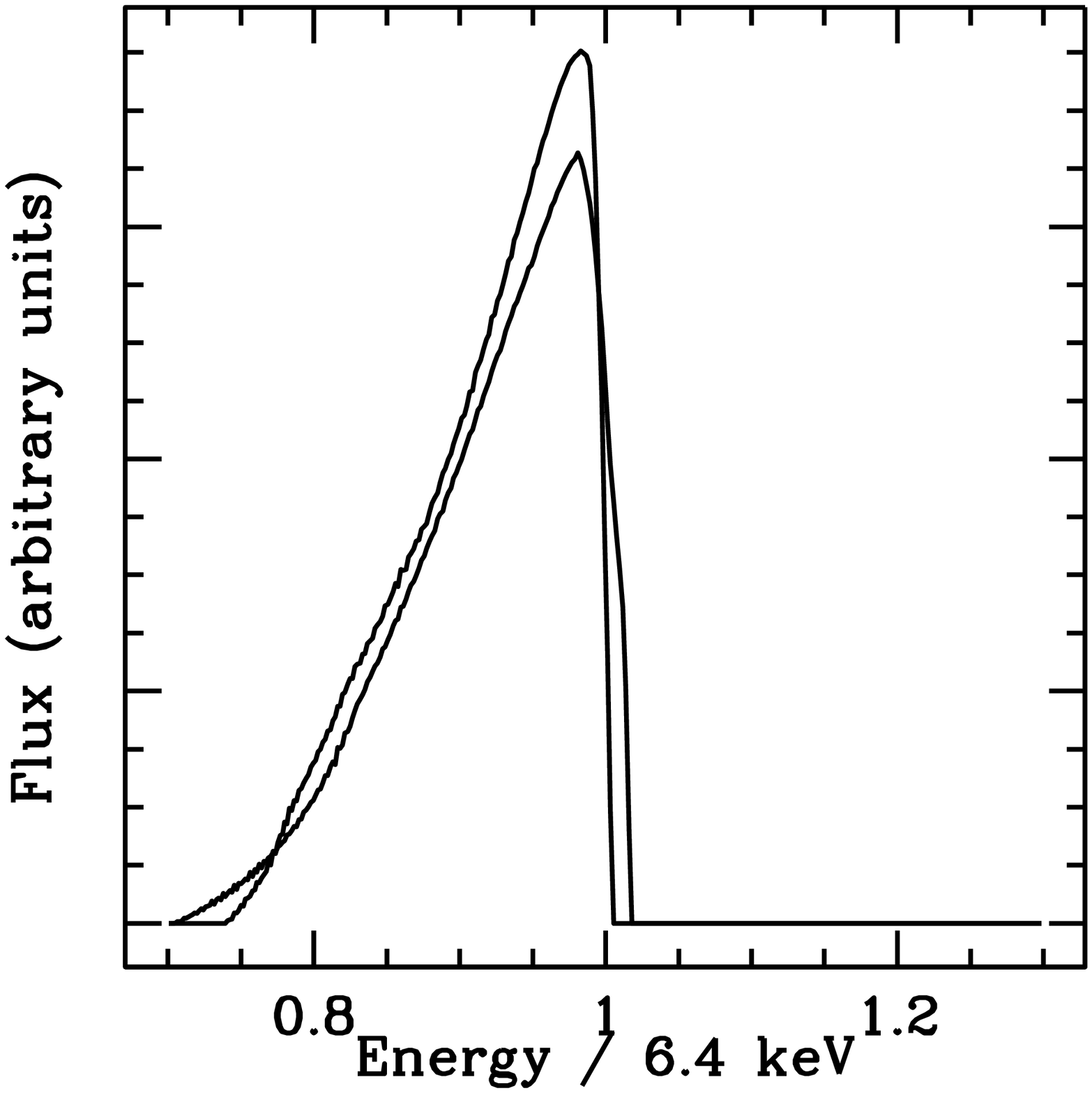,width=1.8in}\\

\noindent\textbf{Figure 4:} Profiles from an ADAF disc with $\epsilon=0.1$.
X-ray bath dominated ($A=1$, $B=100$). Left to right:
non-shadowed and shadowed.
Top to bottom: $g=0$ and $g=1.5$.
Each plot shows profiles for inclinations of 15 and 60 degrees,
plotted to scale (i.e. not separately normalised). The profile
with the higher blue cutoff is at 60 degrees.
\end{figure}

The most notable feature of the profiles from the $\epsilon=10$
ADAF disc, in Fig.5, is that cases with $g=0$ are of virtually
identical form as the standard thin disc profiles, and
show the same inclination dependence as the thin discs! This is
not too surprising because the velocities of equations
\ref{eq:two}-\ref{eq:three} give a magnitude very close the
Keplerian magnitude and the angular motion is more important
than the radial motion.

The profiles with $g=1.5$ are dominated by emission from the central
regions, which results in the large amounts of red and blue shifted
flux due to Doppler and gravitational effects.

\begin{figure}
\epsfig{file=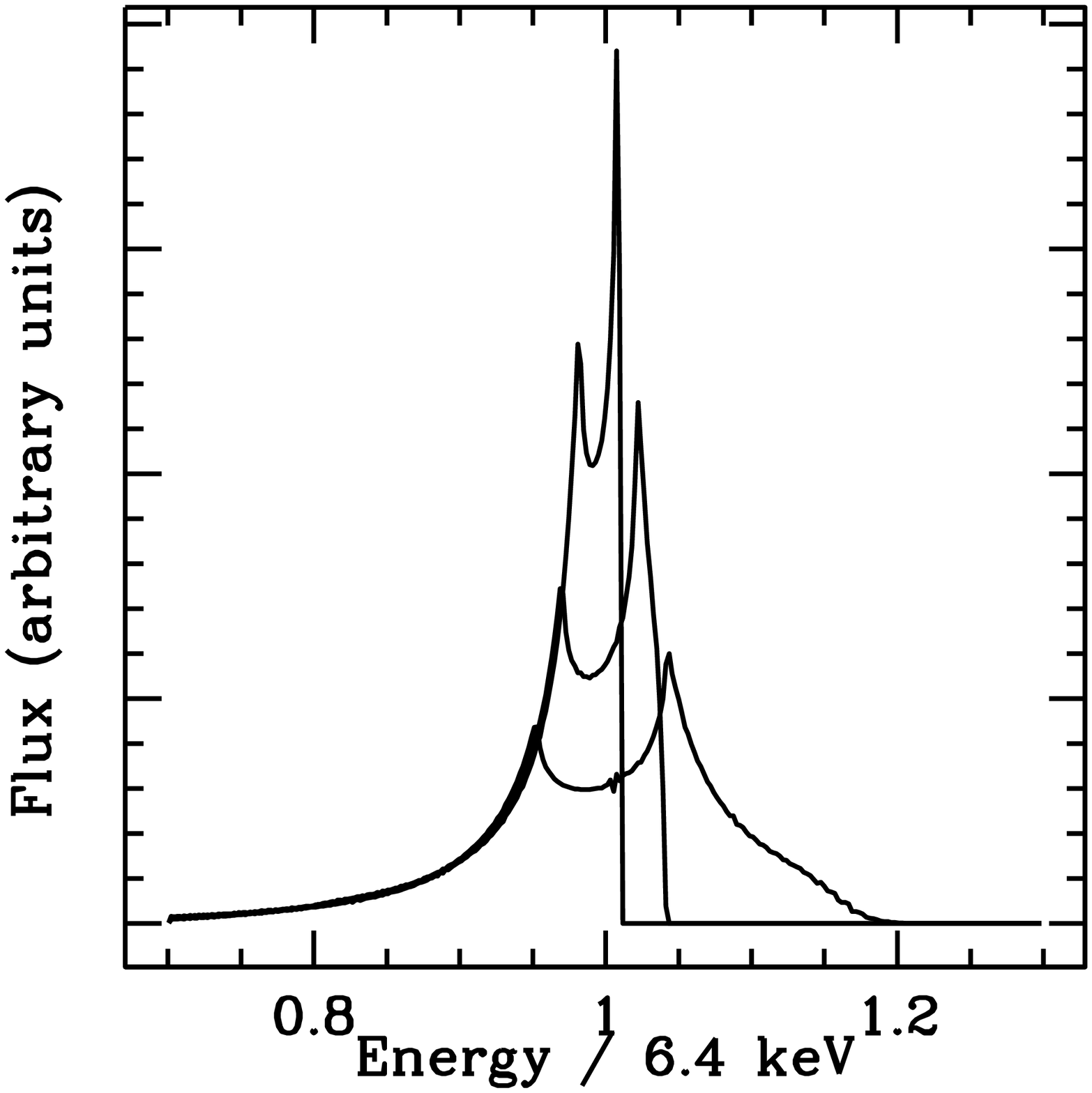,width=1.8in}\epsfig{file=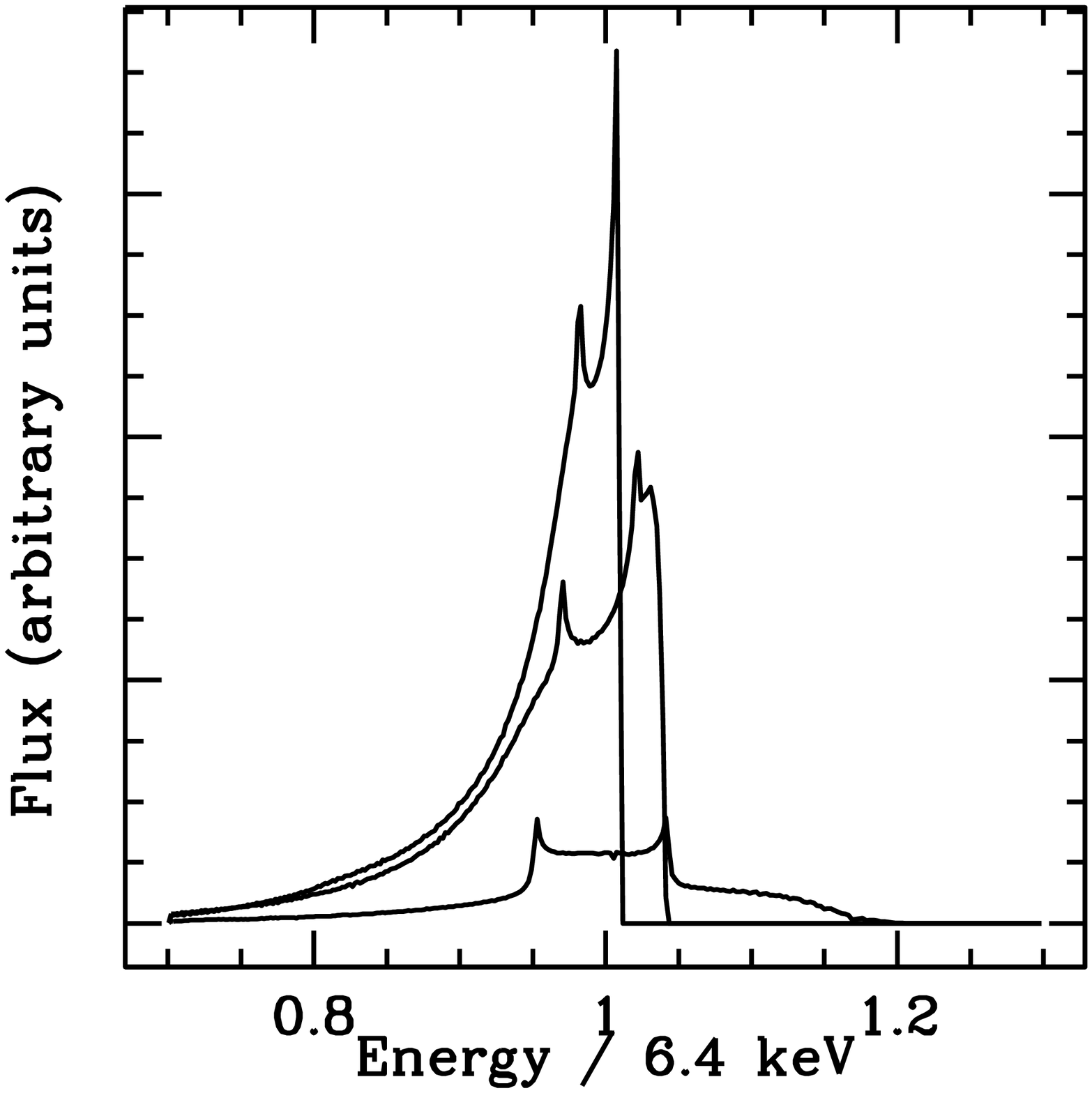,width=1.8in}\\
\epsfig{file=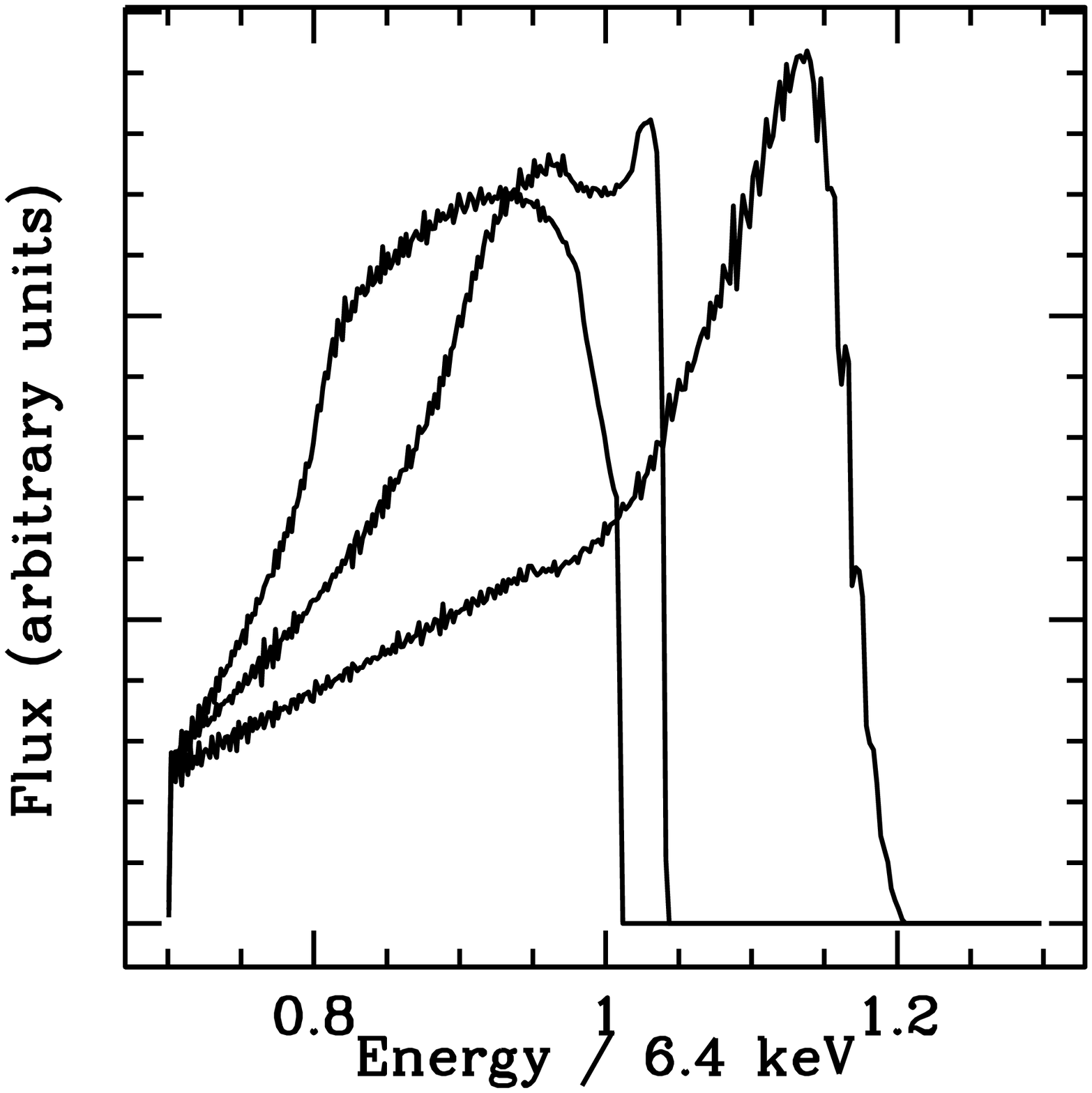,width=1.8in}\epsfig{file=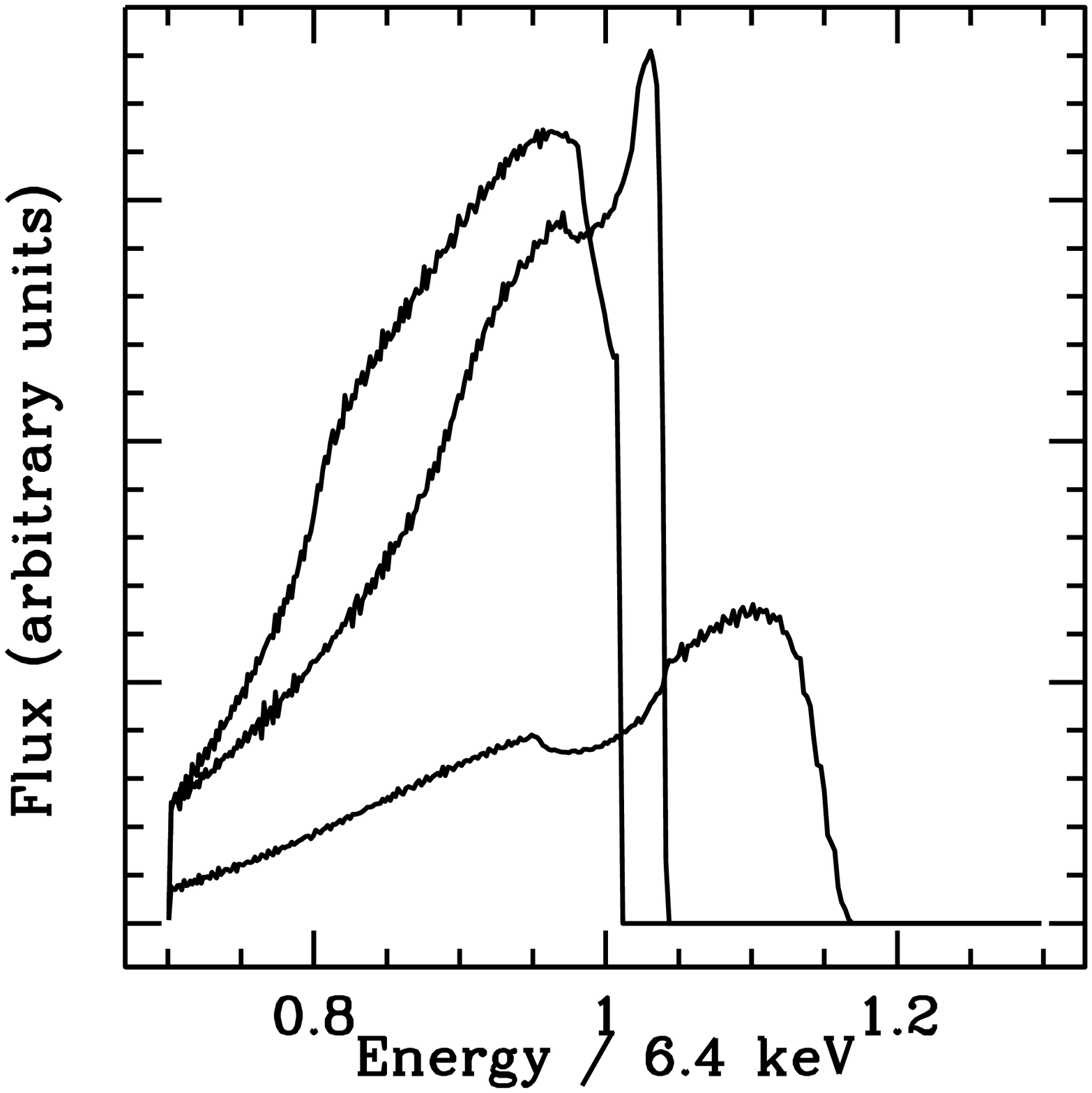,width=1.8in}\\

\noindent\textbf{Figure 5:} Profiles from an ADAF disc with $\epsilon=10$.
Central illumination dominated ($A=10$, $B=1$). Left to right:
non-shadowed and shadowed.
Top to bottom: $g=0$ and $g=1.5$.
Each plot shows profiles for inclinations of 15, 30 and 60 degrees,
plotted to scale (i.e. not separately normalised).
The profiles with higher blue cutoffs are at higher inclinations.
\end{figure}

\subsection{Radial infall profiles}

We consider a radially infalling disc dominated by central
illumination. The profile is independent of inclination because
of the spherical symmetry, so only one inclination angle is considered.
The profiles are shown in Fig.6. To show explicitly the effect
of shadowing, the non-shadowed profiles are plotted to scale
with the shadowed profiles in the left column.

The profiles show a significant amount of blueshifted flux and little
redshifted flux. This is the immediate consequence of the central illumination
which means the blobs falling inwards towards the observer are
visible, whilst those falling inwards away form the observer are
illuminated on the side obscured from the observer. In the case
where the profile is dominated by contributions from the central regions
($g=1.5$), we see a gravitational redshift and more blueshift from
the fast inner flow.

\begin{figure}
\epsfig{file=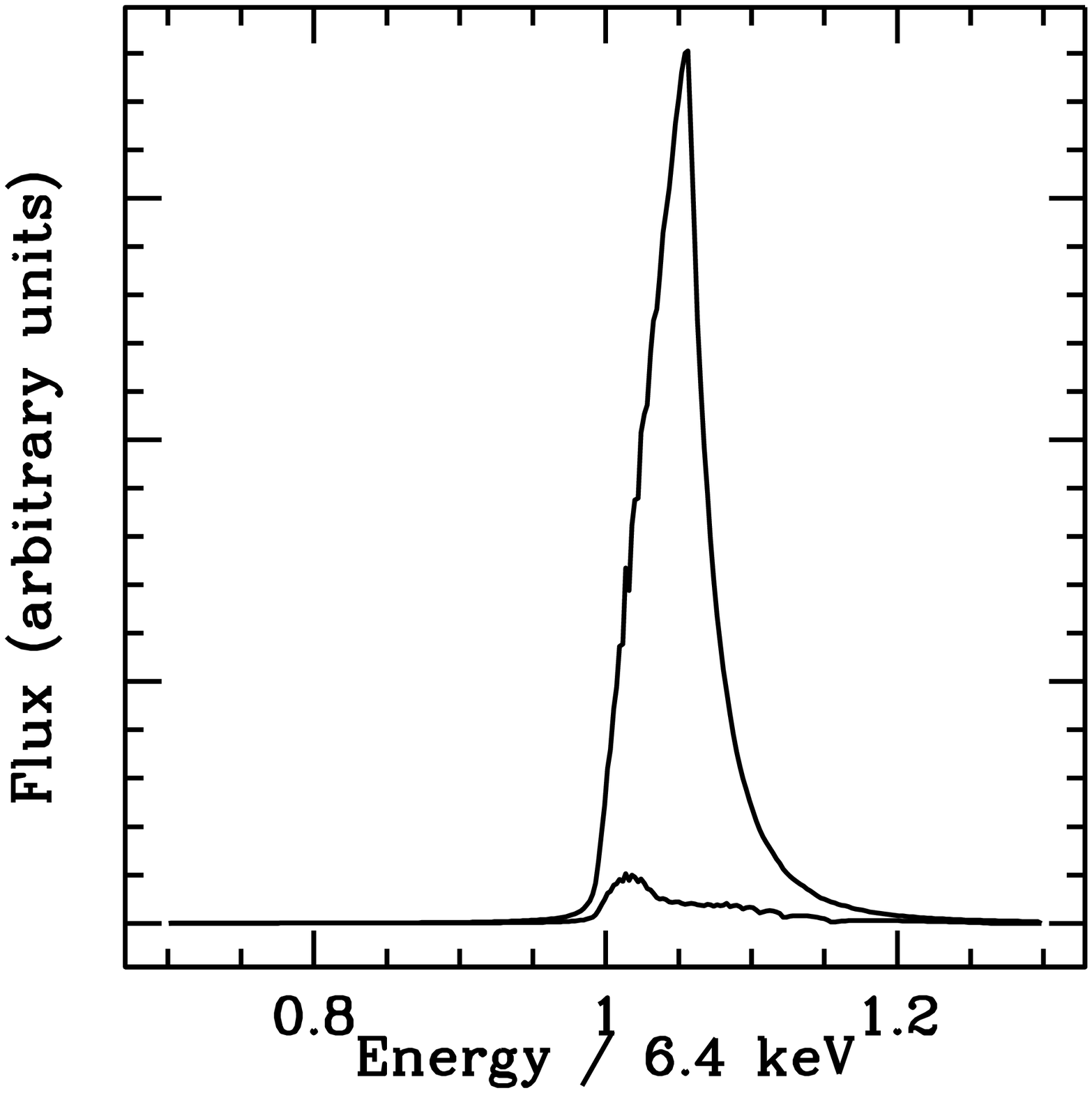,width=1.8in}\epsfig{file=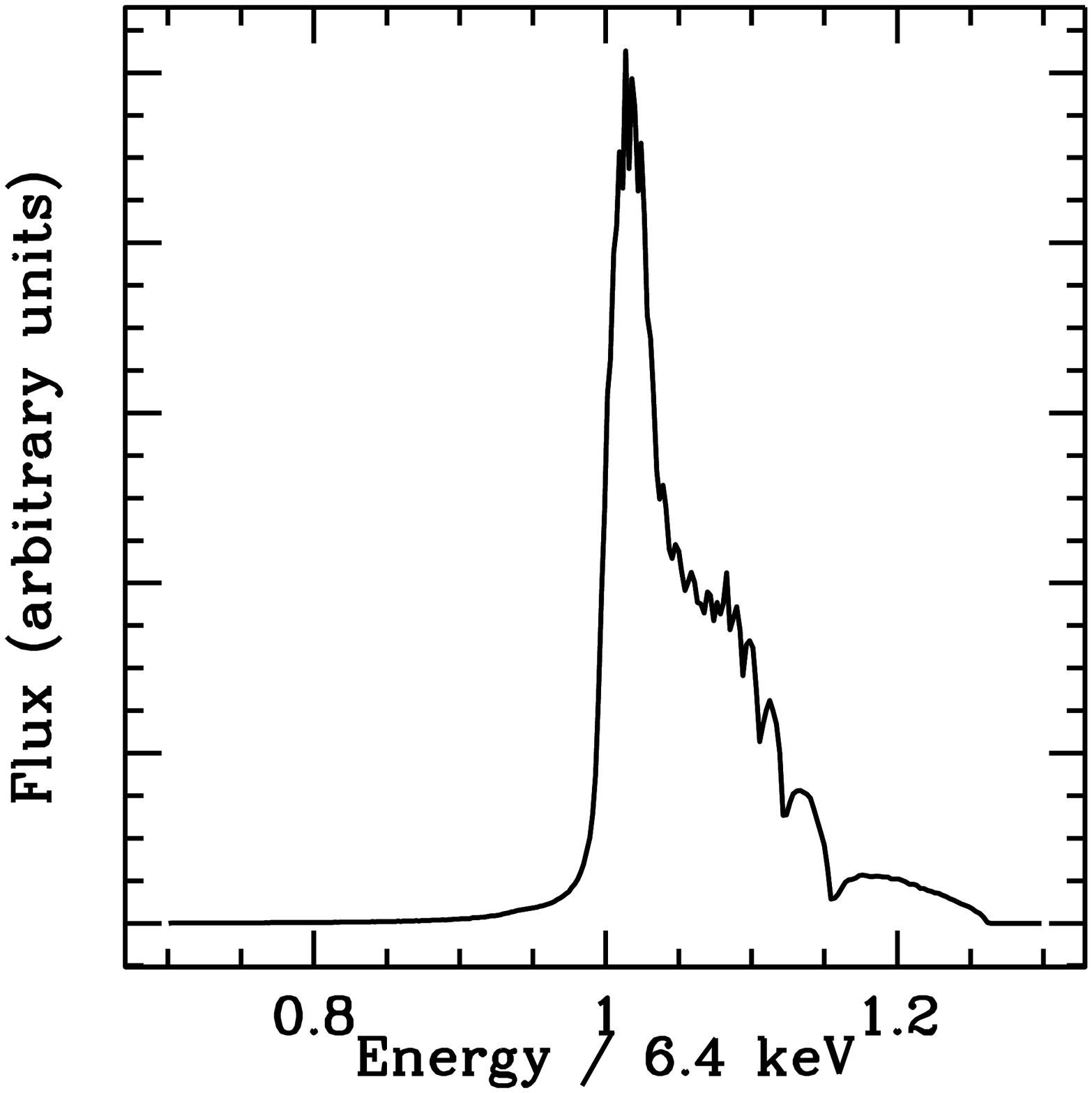,width=1.8in}\\
\epsfig{file=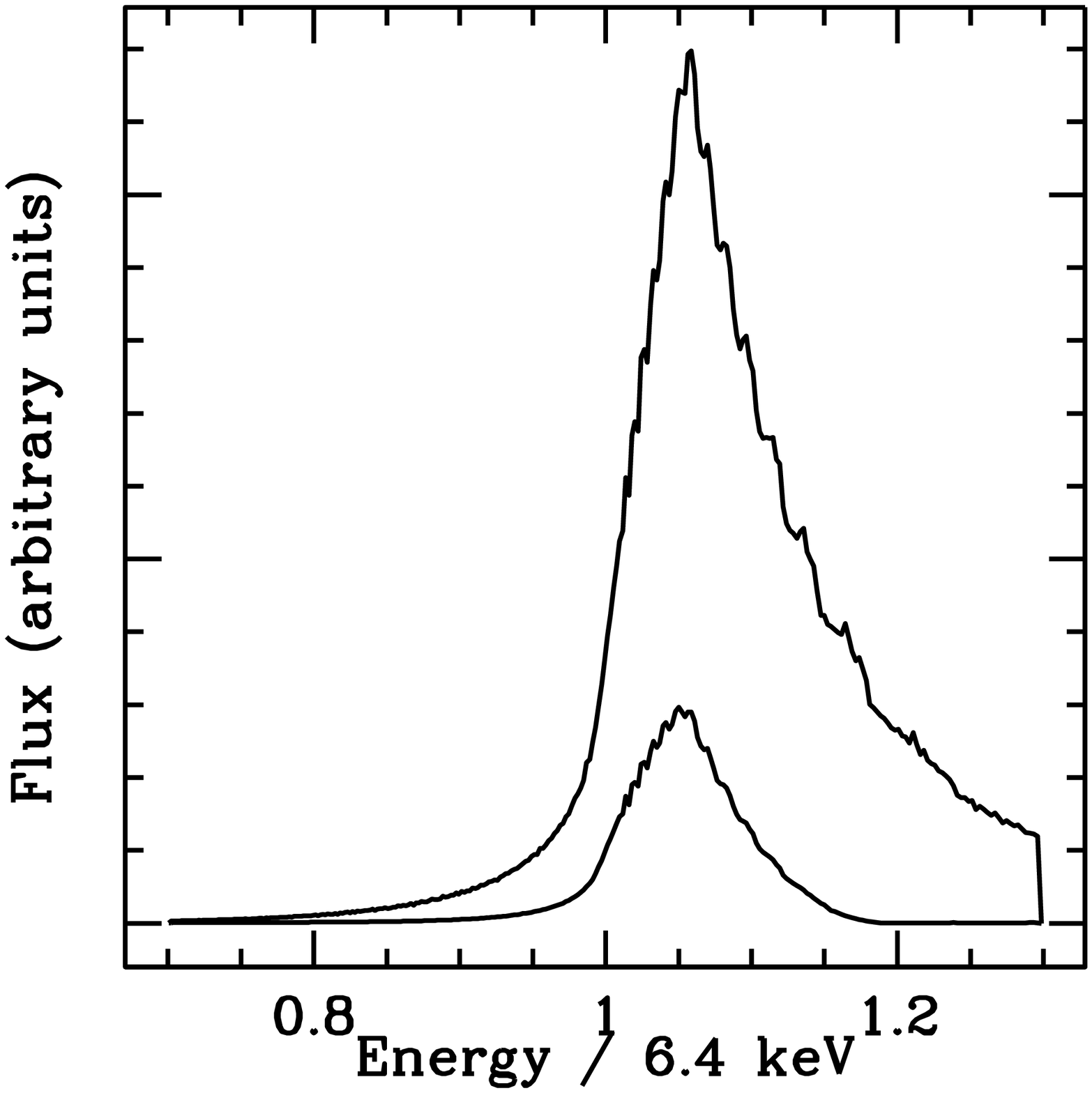,width=1.8in}\epsfig{file=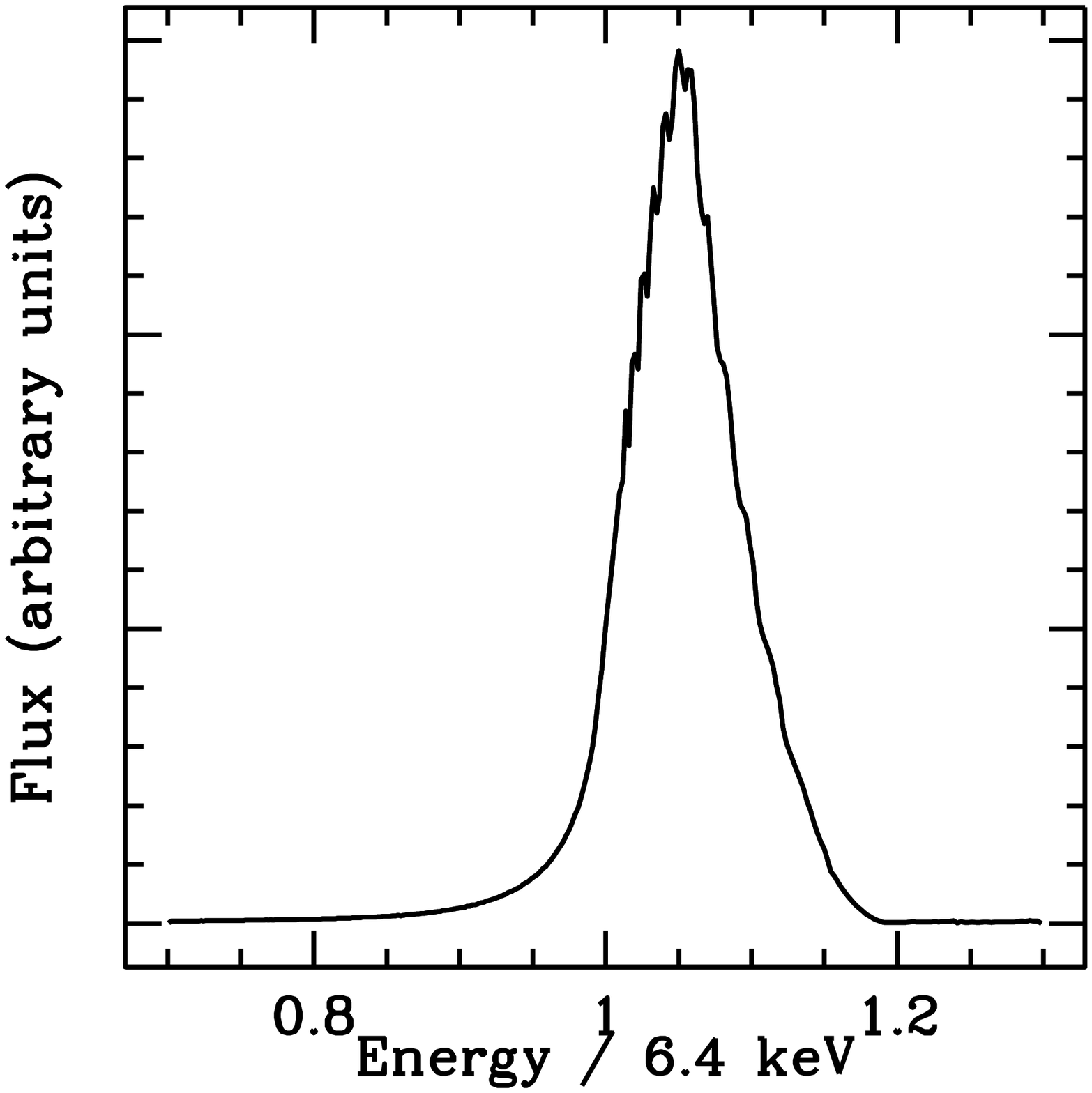,width=1.8in}\\

\noindent\textbf{Figure 6:} Profiles from a radially infalling disc.
Central illumination dominated ($A=10$, $B=1$). Left to right:
non-shadowed+shadowed (to scale) and shadowed (scaled up).
Top to bottom: $g=0$ and $g=1.5$.
\end{figure}

\subsection{Outflow profiles}

We consider a disc with the bipolar outflow solutions of equations
\ref{eq:f} to \ref{eq:s}. The disc has $\epsilon=0.1$ and is
dominated by central illumination.

The profiles are very similar to those generated by the ADAF
solutions without outflows. Here we show the 30 degree inclination profiles,
but the profiles for different inclinations are very similar,
as for the ADAF cases. The double peak for the
outer region dominated case is seen in the top row of Fig. 7.
The redshift dominates the flux in both cases. This is  
because of the gravitational redshift, and because the use of central
illumination means that the observer primarily sees emission from 
the clouds moving away from his/her line of sight. This is 
opposite to the radial infall case.

\begin{figure}
\epsfig{file=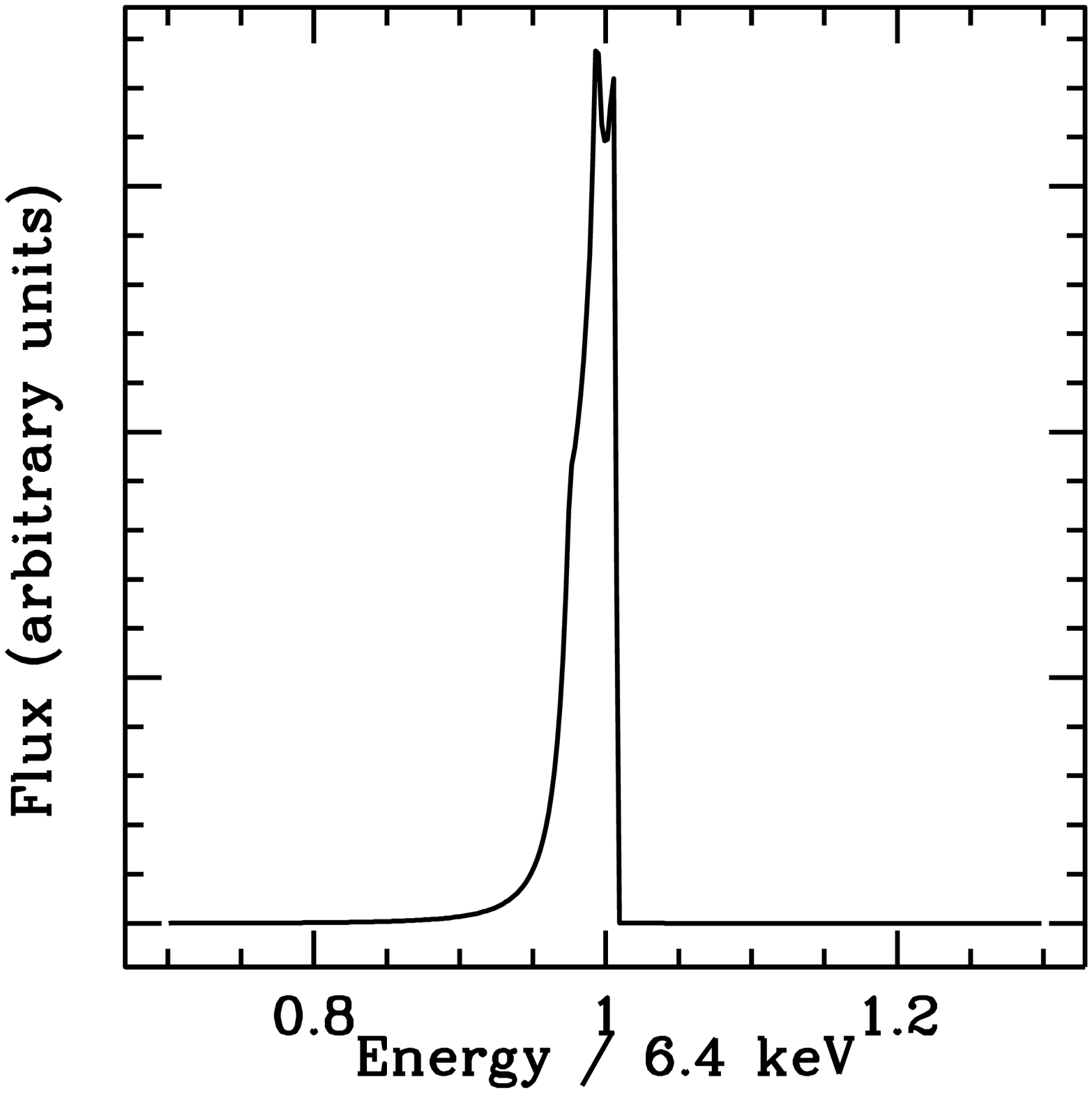,width=1.8in}\epsfig{file=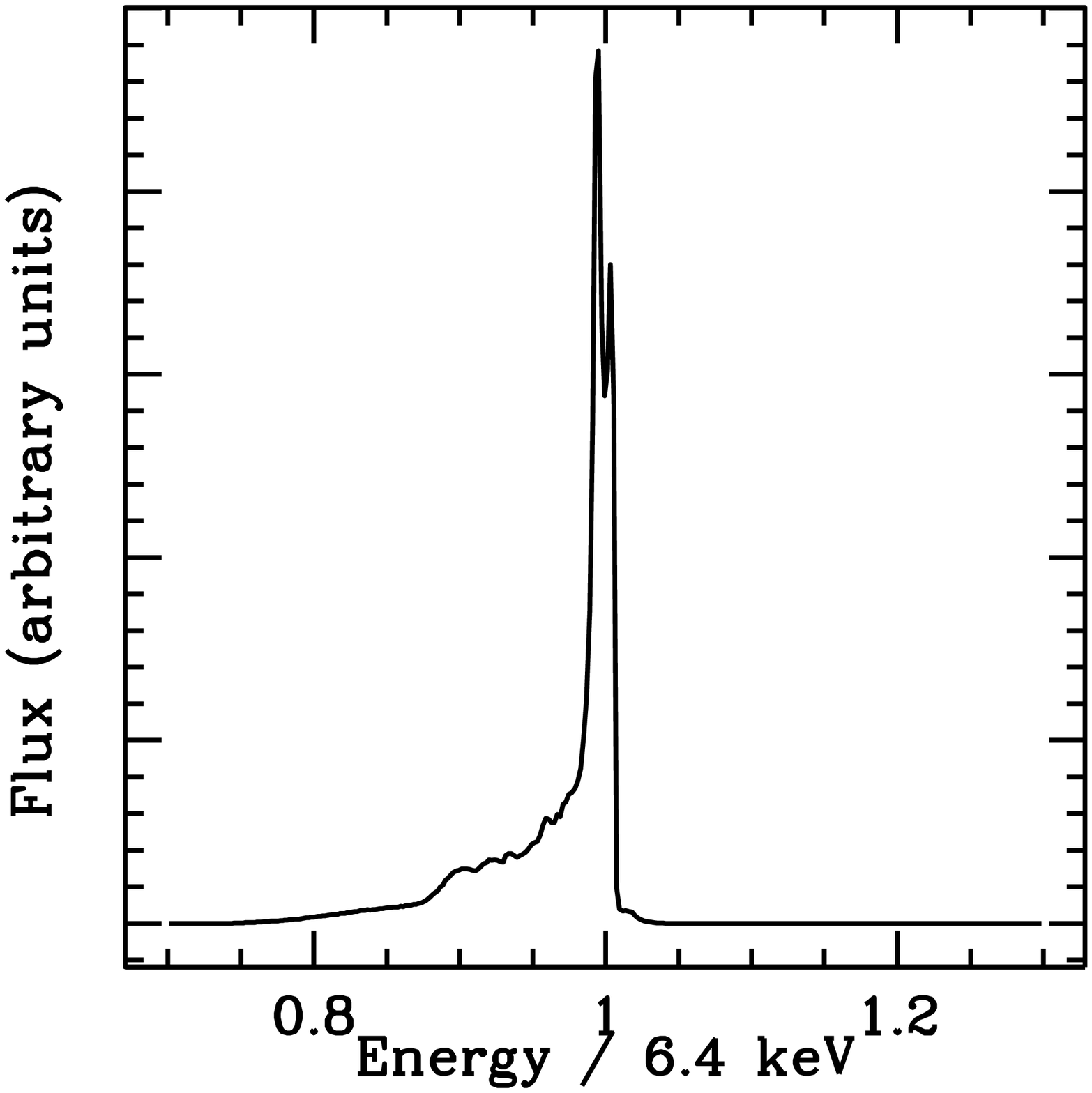,width=1.8in}\\
\epsfig{file=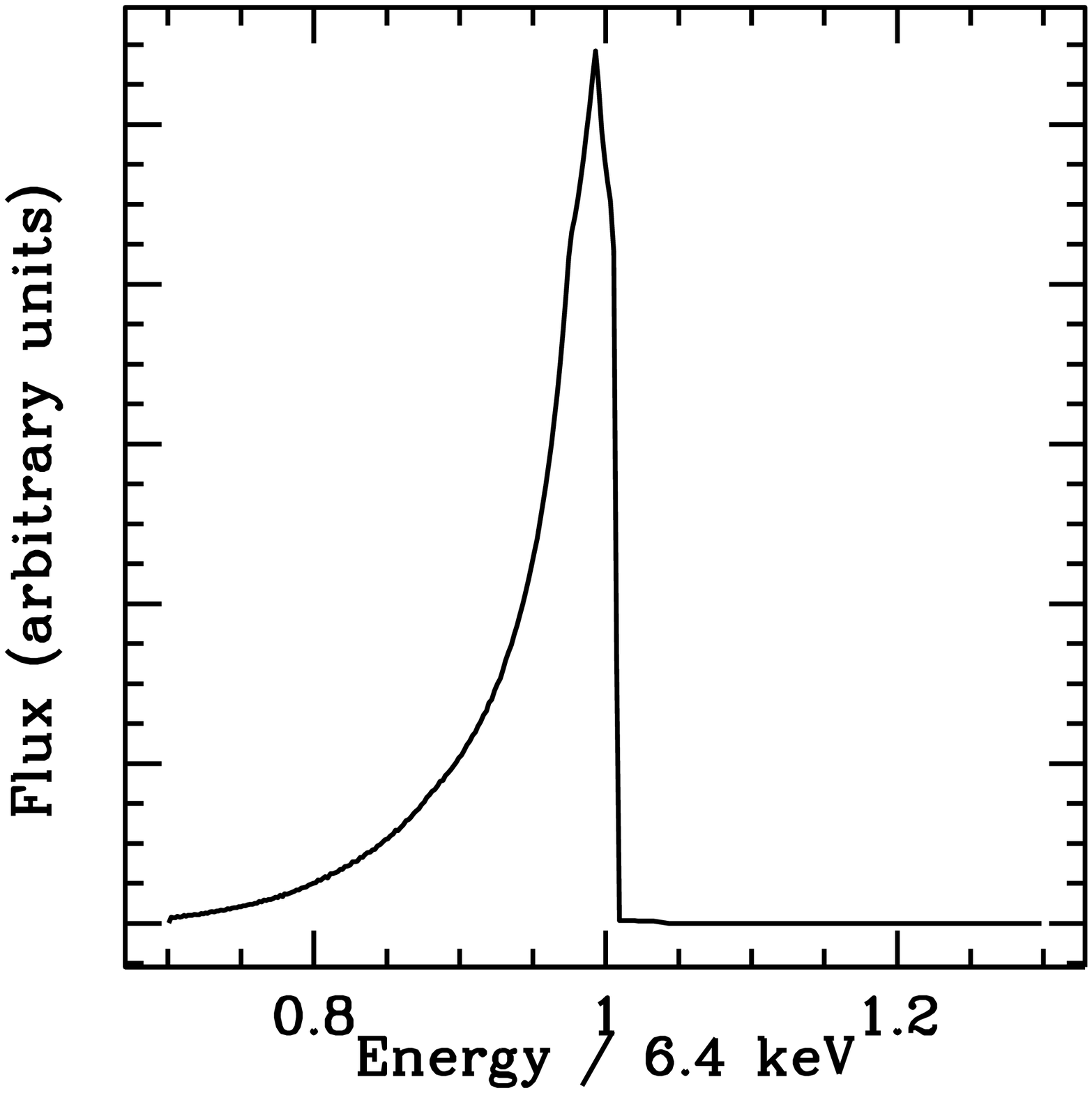,width=1.8in}\epsfig{file=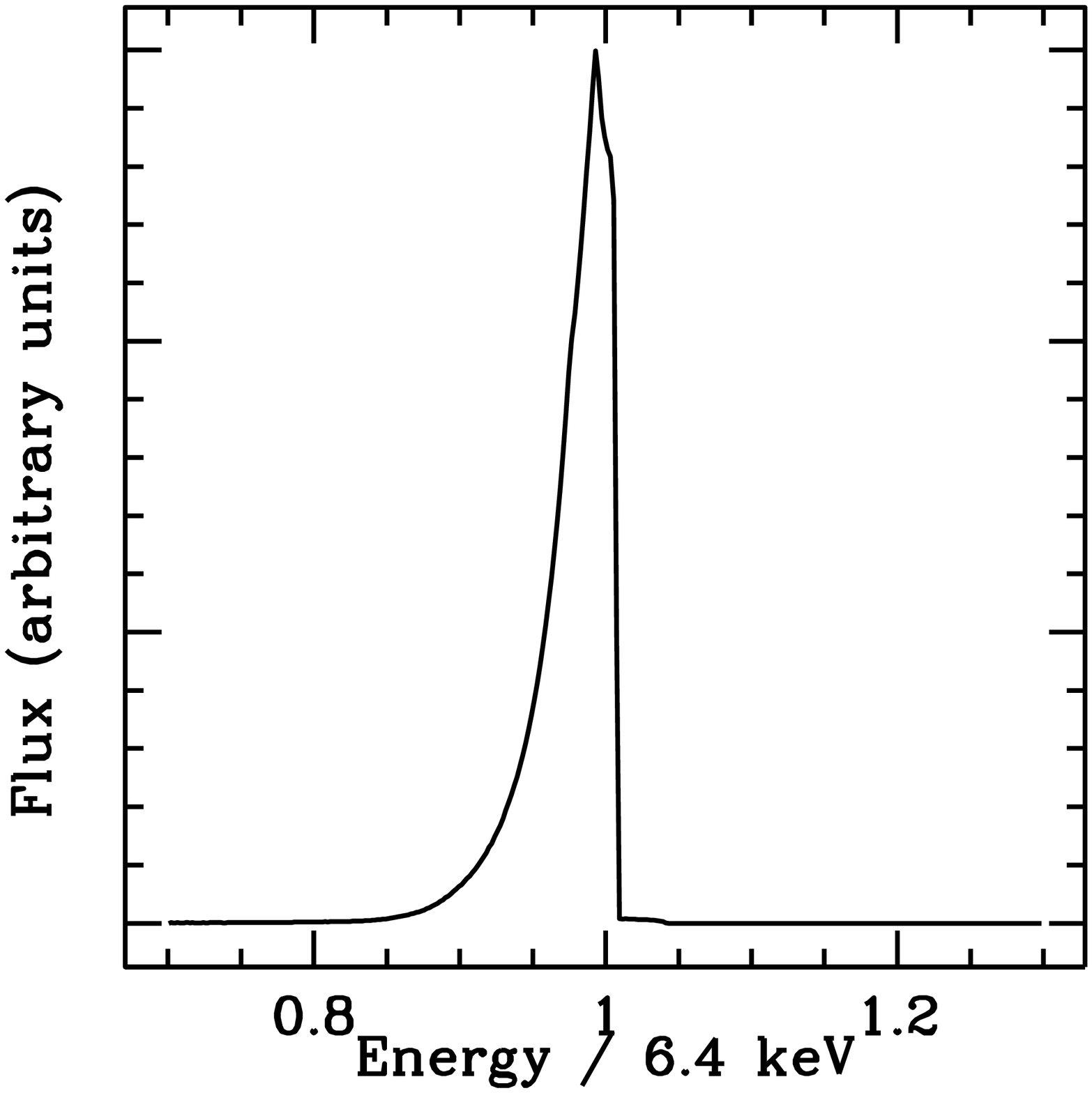,width=1.8in}\\

\noindent\textbf{Figure 7:} Profiles from a bipolar outflow disc.
Central illumination dominated ($A=10$, $B=1$). Left to right:
non-shadowed+shadowed (to scale) and shadowed (scaled up).
Top to bottom: $g=0$ and $g=1.5$.
Inclination of 30 degrees.
\end{figure}

\section{Discussion and conclusion}

We have considered a model in which cold, dense clouds are
embedded in an otherwise optically thin, but
geometrically thick accretion flow. Assuming these clouds
reprocess primary X-ray emission from a central source
and an X-ray bath, we have computed the expected 
fluorescent iron line profiles.
This is a different scenario to the standard calculation
of line profiles from thin, flat discs. 

We considered various solutions for the dynamics of the disc:
ADAFs, pure radial infall models, and bipolar outflows. We allow
these solutions to specify the velocity, number density and size of
the clouds at different points of the engine. In calculating
the line profiles, a new effect compared to the thin disc models
is the obscuration of flux by the clouds themselves,
which are optically thick.

Many of the resulting line profiles are similar to those
from thin disc models, suggesting that thin discs
do not provide a unique model that is able to reproduce
observations of iron line profiles. This suggests
that cloud models warrant further study.
Previous work has focused primarily on the X-ray continuum
(Sivron and Tsuruta 1993; Bond and Matsuoka 1993;
Nandra and George 1994; Collin-Souffrin et al. 1996)
but there has been little  modelling of the relativistic iron
fluorescent line for the cloud model. Karas et al. 2000
considered non-relativistic profiles with large clouds
at one radius.

A further incentive to consider thick disks with embedded clouds is that
they may help resolve problems arising in thin disc modelling
which relate to the inclination of discs and the reprocessed fraction. 
This includes the puzzling preference for low inclinations for  
Seyfert 2s (Turner et al. 1998), some incongruences
between inferred inclinations from iron line
modelling and those determined from other methods
(Nishiura et al. 1998; Sulentic et al. 1998a; Wang et al. 1999), 
and the observed reprocessed fraction appearing to be too low 
and/or uncorrelated with line width (e.g. Lee et al. 2000; Chiang et al. 2000;
Weaver 2000; Sulentic et al. 1998b).
In this regard, 
we find that many of our  embedded cloud, thick disc 
profiles (modelled with ADAF type flows)
are relatively insensitive to the
inclination angle because a significant pressure support
means that the bulk velocities are smaller than the Keplerian values.

This latter feature also means that many of our 
profiles have sharp blue cut-offs because they are dominated by the  
gravitational redshift. Whether the sharp blue cutoff
remains true in the Kerr case, where the disc
extends closer to central hole and hence can acquire greater
velocities should be investigated. Note that this will also depend
fundamentally on the innermost radius at which the reprocessing
clouds can survive. A thick geometry can also allow a wider range of
reprocessed fractions, depending on the cloud distribution
and amount of obscuration. 

To summarise, the embedded cloud+ thick disc model allows a wider range of
disc flow solutions to show reprocessed X-rays lines, 
and highlights how an alternative geometry  can produce profiles similar
to those from thin discs.  The model can also lead to profiles 
with different properties to those of the thin discs,
such as less sensitivity to inclination angle. As such,
we suggest the cloud model deserves further investigation.
This would include more consideration of the mechanisms of 
cloud formation and sizes,
generalisation to the Kerr metric, and application to specific sources.

\section*{Acknowledgements} Thanks to Clovis Peres for 
many fruitful discussions and early collaborations on this subject.  
SAH acknowledges the hospitality of the University of Rochester and the
REU program. EB acknowledges partial 
support from DOE grant DE-FG02-00ER54600.

\end{document}